\documentclass{UApreprint}
\usepackage{epsfig}
\usepackage{graphics}
\usepackage{fullpage}
\usepackage{verbatim}
\usepackage{feynmp}
\usepackage[tight]{subfigure}
\newcommand{\be}{\begin{equation}}
\newcommand{\ee}{\end{equation}}
\newcommand{\bea}{\begin{eqnarray}}
\newcommand{\eea}{\end{eqnarray}}
\newcommand{\bean}{\begin{eqnarray*}}
\newcommand{\eean}{\end{eqnarray*}}

\begin{document}
%\draft

\title{Constraints on Scalar Couplings from $\pi^\pm \rightarrow l^\pm\nu_l$}
\author{Bruce A. Campbell and David W. Maybury}
\address{Department of Physics, University of Alberta, Edmonton  AB  T6G 2J1,
CANADA}
\date{}

\abstract{New interactions with Lorentz scalar structure, arising from physics beyond the standard model of electroweak interactions, will
induce effective pseudoscalar interactions after renormalization by weak interaction loop corrections. Such induced pseudoscalar interactions
are strongly constrained by data on $\pi^\pm \rightarrow l^\pm \nu_l$ decay. These limits on induced pseudoscalar interactions imply limits
on the underlying fundamental scalar interactions that in many cases are substantially stronger than limits on scalar interactions from direct
$\beta$-decay searches.
}

\archive{hep-ph/0303046}
\preprintone{ALTA-TH-06-03}
\preprinttwo{}

\submit{}

\maketitle

%%%%%%%%%%%%%%%%%%%%%%%%%%%%%%%%%%%%%%%%%%%%%%%%%%%%%%%%%%%%%%%%%%
\section{Introduction}

While there is strong support for the $V-A$ form of the charged weak current,
it is possible that new physics at or above the weak scale could give rise to scalar interactions that would compete
with standard model
processes. Examples of such possible physics include the exchange of extra Higgs multiplets which could enter
the theory at scales from the
Z mass upwards \cite{PDG}, leptoquarks which could be present at scales above 200 GeV \cite{PDG},
contact interactions from quark/lepton compositeness which could be present at the TeV scale
\cite{PDG}, or strong gravitational interactions in TeV brane world models \cite{PDG}.
Recently, precision experiments \cite{Garcia:1999da,Adelberger:1999ud,Adelberger:1993wq} have searched for
scalar interactions in $\beta$-decay, however, direct experimental constraints on scalar couplings still remain
relatively weak as compared to the corresponding limits on pseudoscalar couplings \cite{PDG,Towner:1995za}.

The precision of the limits on pseudoscalar couplings comes in
part from the fact that the pion, a pseudoscalar meson, has a
chirally suppressed decay $\pi^\pm \rightarrow l^\pm \nu_l$ which
would be sensitive to new pseudoscalar interactions \cite{Bryman}.
These pseudoscalar interactions would be detected by the failure
of the standard model prediction \cite{Marciano:1993sh} for the
chiral suppression in the ratio of branching ratios
$\textstyle{\frac{\Gamma(\pi^- \rightarrow e \bar
\nu)}{\Gamma(\pi^- \rightarrow \mu \bar \nu)}}$. It is the large
chiral suppression factor, by the square of the electron-muon mass
ratio, that allows such a powerful test of new physics that
violates chirality and parity.

In the standard model, the leading contribution to pion decay occurs
through tree level W exchange. At the quark level, this is the same process that is involved
in the $\beta$-decay of a nucleon ignoring the spectator quarks.
While the pion cannot decay through a scalar interaction,
the pion can decay through induced pseudoscalar interactions generated from the electroweak renormalization of the scalar
couplings. It is of considerable interest to use limits on the induced pseudoscalar couplings to set indirect limits on the size of the
underlying scalar interactions.
%Br($e\nu/\mu\nu$).
%$\textstyle{\frac{\Gamma(\pi^- \rightarrow e \bar \nu)}{\Gamma(\pi^- \rightarrow \mu \bar \nu)}}$.

In the following sections we outline our methods and estimate the limits on the size of scalar couplings based
on the indirect effects from charged pion decay. We use
general operator techniques to obtain model independent results and we combine these results with data from pion decay
and also muon capture, to constrain the scalar couplings indirectly. We also discuss some of the implications of these
results and comment on prospects for future searches for scalar interactions.

\section{Pion Physics and New Pseudoscalar Interactions}
\label{pionp}
Consider constructing an effective Lagrangian and matrix element for the process $\pi^\pm \rightarrow l^\pm \nu_l$
in the presence of pseudoscalar interactions. We can set limits on the strength of the pseudoscalar interactions from their interference with tree
level W exchange. Since the pion is a pseudoscalar, we can use the following
relations for current matrix elements,
\bea
\label{onshellst}
&&\left<0\left|\bar u \gamma_\mu \gamma_5 d\right| \pi(p) \right> = i \sqrt{2}
f_\pi p_\mu \nonumber \\
&& \left<0\left|\bar u \gamma_5 d\right| \pi(p) \right> = i \sqrt{2} \tilde
f_\pi = i \sqrt{2} \hspace{1mm} \frac{f_\pi m^2_\pi}{m_u + m_d} \nonumber \\
&& \left<0\left|\bar u \sigma^{\mu \nu}\gamma_5 d\right| \pi(p) \right> =0
\nonumber \\
&& \left<0\left|\bar u \sigma^{\mu \nu} d\right| \pi(p) \right> =0,
\eea
where $f_\pi = 93 \hspace{1mm}\textnormal{MeV}$ and
$\tilde f_\pi = 1.8 \times 10^5 \hspace{1mm}\textnormal{MeV}^2$.
The matrix element for the tree level W contribution can easily be constructed by using eq.(\ref{onshellst}),
giving;
\be
\mathcal{M}_{W^\pm} = G_F f_\pi \cos \theta_c [\bar l \gamma^\mu(1-\gamma_5)
\nu_l] p_\mu,
\ee
where $p_\mu$ is the pion momentum and $\theta_c$ is the Cabibbo angle.
A pseudoscalar contribution with left-handed neutrinos in the final state
can be expressed as a four-fermi contact operator,
\be
\mathcal{L}_P = -i \frac{\rho}{2 \Lambda^2}[\bar l (1-\gamma_5)\nu_l][\bar u \gamma_5 d]
\ee
where $\rho$ is the pseudoscalar coupling constant. This expression can be
converted to a matrix element using eq.(\ref{onshellst}),
\be
\mathcal{M}_P = \rho \frac{\tilde f_\pi}{\sqrt{2} \Lambda^2}[\bar l (1-\gamma_5) \nu_l].
\ee
In the presence of a pseudoscalar interaction, the overall matrix element for the process $\pi^\pm \rightarrow l^\pm \nu_l$ is the coherent sum,
$\mathcal{M}_{P}+ \mathcal{M}_{W^\pm}=\mathcal{M}_l$.
\be
\mathcal{M}_l= G_F f_\pi \cos \theta_c[\bar l \gamma^\mu(1-\gamma_5)\nu_l]p_\mu + \frac{\rho \tilde
f_\pi}{\sqrt{2} \Lambda^2}[\bar l (1-\gamma_5)\nu_l]
\ee
Having constructed the matrix element, we can now estimate the ratio of branching
ratios,
\be
\frac{\Gamma(\pi^- \rightarrow e \nu_e)}{\Gamma(\pi^- \rightarrow \mu \nu_\mu)} =
\frac{(m_\pi^2 -m_e^2)}{(m_\pi^2-m_\mu^2)} \frac{\left<|M_{e
\nu}|^2\right>}{\left<|M_{\mu\nu}|^2\right>}.
\ee
Summing over final states of the squared matrix element we have
\bea
\left<|\mathcal{M}_l|^2\right> &=& 4 \hspace{.5mm} G_f^2 f_\pi^2 \cos^2 \theta_c m_l^2(m_\pi^2-m_l^2)
+ 8 \frac{ G_F \tilde f_\pi f_\pi \cos \theta_c \rho}{\sqrt{2} \Lambda^2}m_l(m_\pi^2 -m_l^2) \nonumber \\
&&+ 2\frac{\rho^2 \tilde f_\pi^2}{\Lambda^4} (m_\pi^2-m_l^2).
\eea
For simplicity we have assumed that the pseudoscalar coupling is
real, however, in general $\rho$ may be complex. The more general expression is obtained by making
the following replacements,
\bea
\rho &\rightarrow& \frac{\rho+\rho^*}{2} = \textnormal{Re}(\rho) \nonumber \\
(\rho)^2 &\rightarrow& |\rho|^2.
\eea
We find that the branching ratio is given by
\be
\label{br}
\frac{\Gamma(\pi^- \rightarrow e \nu_e)}{\Gamma(\pi^- \rightarrow \mu \nu_\mu)}
=\frac{(m_\pi^2 -m_e^2)}{(m_\pi^2-m_\mu^2)} \left[\frac{m_e^2(m_\pi^2-m_e^2)
+R_e}{m_\mu^2(m_\pi^2-m_\mu^2) + R_\mu} \right],
\ee
where the $R_{e,\mu}$ functions are
\be
\label{Semu}
R_{e,\mu} = \sqrt{2} \frac{\tilde f_\pi \mathrm{Re}(\rho)}{G_F f_\pi \Lambda^2 \cos \theta_c} m_{e,\mu}(m_\pi^2
-m_{e,\mu}^2) + \frac{|\rho|^2 \tilde f_\pi^2}{2 f_\pi^2 G_F^2 \Lambda^4 \cos^2 \theta_c}(m_\pi^2-m_{e,\mu}^2).
\ee

Thus far we have only discussed interactions with left-handed
neutrinos in the final state. The inclusion of right-handed
neutrinos requires a modification since pseudoscalar contributions
to decays with right-handed neutrinos in the final state cannot
interfere with the W exchange graph; hence the contributions to
the rate add incoherently. With right-handed neutrinos, the
expression for the matrix element becomes,
\be \mathcal{M}_P=
\frac{\rho^\prime \tilde f_\pi}{\sqrt{2} \Lambda^2} [\bar l
(1+\gamma_5)\nu_l], \ee where $\rho^\prime$ is the pseudoscalar
coupling involving right-handed neutrinos. Defining \be T \equiv
\frac{(m_\pi^2 -m_e^2)^2}{(m_\pi^2-m_\mu^2)^2}
\frac{m_e^2}{m_\mu^2} = 1.28 \space \times \space 10^{-4},
\ee
we can express the branching ratio as
\bea
\label{BRemu} \frac{\Gamma
(\pi^- \rightarrow e \nu_e)}{\Gamma (\pi^- \rightarrow \mu
\nu_\mu)}&=& T \left(\frac{1 + \sqrt{2}\frac{\tilde f_\pi
\mathrm{Re}(\rho_e)}{G_F \Lambda^2 f_\pi \cos \theta_c m_e} +
\frac{|\rho_e|^2 \tilde f_\pi^2} {2 G_F^2 \Lambda^4 f_\pi^2 \cos^2
\theta_c m_{e}^2} + \frac{|\rho_e^\prime|^{2} \tilde
f_\pi^2}{2 G_F^2 f_\pi^2 \Lambda^4 \cos^2 \theta_c m_e^2}}{1 +
\sqrt{2}\frac{\tilde f_\pi \mathrm{Re}(\rho_\mu)}{G_F \Lambda^2
f_\pi \cos \theta_c m_\mu} + \frac{|\rho_\mu|^2 \tilde f_\pi^2} {2
G_F^2 \Lambda^4 f_\pi^2 \cos^2 \theta_c m_{\mu}^2} +
\frac{|\rho_\mu^\prime|^2 \tilde f_\pi^2}{2 G_F^2 \Lambda^4
f_\pi^2\cos^2 \theta_c m_\mu^2}} \right)
\eea
If we assume either universal scalar couplings or else scalar couplings involving only
the first generation, we obtain the following approximation for
the ratio of decay widths,
\bea
\label{BRa} \frac{\Gamma (\pi^-
\rightarrow e \nu_e)}{\Gamma (\pi^- \rightarrow \mu
\nu_\mu)}&\approx& T \left(1 + \sqrt{2}\frac{\tilde f_\pi
\mathrm{Re}(\rho)}{G_F \Lambda^2 f_\pi \cos \theta_c m_e} +
\frac{|\rho|^2 \tilde f_\pi^2} {2 G_F^2 \Lambda^4 f_\pi^2 \cos^2
\theta_c m_{e}^2} + \frac{|\rho^\prime|^{2} \tilde f_\pi^2}{2
G_F^2 \Lambda^4 f_\pi^2 \cos^2 \theta_c m_e^2} \right) \nonumber \\
&&
\eea
We will discuss the effects of more general generation
dependence of the scalar couplings in section \ref{lasts}. The
theoretical standard model calculation including radiative
corrections is $\textnormal{Br}_\textnormal{th}=(1.2352 \pm
.0005)\times 10^{-4}$ \cite{Marciano:1993sh} and the measured
experimental branching ratio is
$\textnormal{Br}_\textnormal{exp}=(1.230 \pm .0040) \times
10^{-4}$ \cite{PDG,Britton1,Britton2,Czapek1}. Combining the
experimental and theoretical uncertainties in quadrature, we can obtain a bound on
the pseudoscalar couplings at $2\sigma$,
\be
-1.0 \times 10^{-2} \leq \sqrt{2}\frac{\tilde f_\pi
\mathrm{Re}(\rho)}{G_F \Lambda^2 f_\pi \cos \theta_c m_e} +
\frac{|\rho|^2 \tilde f_\pi^2} {2 G_F^2 \Lambda^4 f_\pi^2 \cos^2
\theta_c  m_{e}^2} + \frac{|\rho^\prime|^{2} \tilde f_\pi^2} {2
G_F^2 \Lambda^4 f_\pi^2 \cos^2 \theta_c m_e^2} \leq 2.2
\space \times \space 10^{-3}.
\label{csteq}
\ee

%%%%%%%%%%%%%%%%%%%%%%%%%%%%%%%%%%%%%%%%%%%%%%%%%%%%%%%%%%%%%%%%%
%%%%%%%%%%%%%%%%%%%%%%%%%%%%%%%%%%%%%%%%%%%%%%%%%%%%%%%%%%%%%%%%
\section{Local Scalar Operator Analysis}
\label{opa}
Electroweak interactions can radiatively induce pseudoscalar operators from pure
scalar interactions.
Suppose that at some scale $\Lambda$ there exists new physics that generates a purely scalar four-fermi
interaction. It may be due to the exchange of fundamental scalars or it may be due to a
variety of other physics such as compositeness, extra dimensions, leptoquarks, et cetera.
Independent of the details of the new physics that generates the scalar interactions, they
will appear as non-renormalizable four-fermi scalar contact operators below the scale $\Lambda$.

In order to facilitate power counting, the $\overline{\textnormal{MS}}$ scheme is most often
used with effective field theory \cite{Manohar:1996cq}. The $\overline{\textnormal{MS}}$ scheme
(or any mass independent subtraction scheme) presents the subtlety that heavy
particles do not decouple in beta function calculations. That is, mass
independent renormalization schemes do not satisfy the conditions of the
Applequist-Carazzone theorem \cite{Manohar:1996cq}. This is dealt with by simply integrating
out the heavy fields by hand at their associated scale. Thus whether we analyze the
effective interactions in a UV complete theory or in the effective theory, we will arrive at the same
renormalization group running
(up to threshold corrections) provided that we are only interested in results below
$\Lambda$ and only up to some finite power of $(\frac{1}{\Lambda})$.

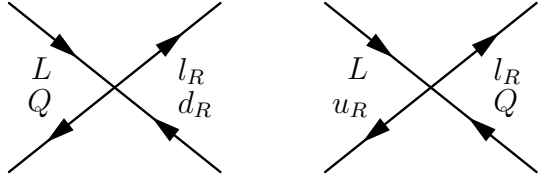
\begin{figure}
    \begin{center}
    \begin{fmffile}{corr1}
    \begin{fmfgraph*}(80,80)
        \fmfbottom{i1,i2}
        \fmftop{o1,o2}
        \fmf{fermion,label=$Q$,label.side=right}{v1,i1}
        \fmf{fermion,label=$d_R$,label.side=right}{i2,v1}
        \fmf{fermion,label=$L$,label.side=right}{o1,v1}
        \fmf{fermion,label=$l_R$,label.side=right}{v1,o2}
    \end{fmfgraph*}
    \end{fmffile}
\makebox(20,20){}
    \begin{fmffile}{corr2}
    \begin{fmfgraph*}(80,80)
        \fmfbottom{i1,i2}
        \fmftop{o1,o2}
        \fmf{fermion,label=$u_R$,label.side=right}{v1,i1}
        \fmf{fermion,label=$Q$,label.side=right}{i2,v1}
        \fmf{fermion,label=$L$,label.side=right}{o1,v1}
        \fmf{fermion,label=$l_R$,label.side=right}{v1,o2}
    \end{fmfgraph*}
    \end{fmffile}
    \caption{$\mathcal{O}_1$ and $\mathcal{O}_2$, Type A contact interactions
    \label{fig3}}
    \end{center}
\end{figure}

\begin{figure}
    \begin{center}
    \begin{fmffile}{corr3}
    \begin{fmfgraph*}(80,80)
        \fmfbottom{i1,i2}
        \fmftop{o1,o2}
        \fmf{fermion,label=$Q$,label.side=right}{v1,i1}
        \fmf{fermion,label=$d_R$,label.side=right}{i2,v2}
        \fmf{fermion,label=$L$,label.side=right}{o1,v1}
        \fmf{fermion}{v2,v1}
        \fmf{fermion}{v1,v3}
        \fmf{fermion,label=$l_R$,label.side=right}{v3,o2}
        \fmffreeze
        \fmf{boson}{v3,v2}
    \end{fmfgraph*}
    \end{fmffile}
\makebox(20,20){}
    \begin{fmffile}{corr4}
    \begin{fmfgraph*}(80,80)
        \fmfbottom{i1,i2}
        \fmftop{o1,o2}
        \fmf{fermion,label=$Q$,label.side=right}{v1,i1}
        \fmf{fermion,label=$d_R$,label.side=right}{i2,v2}
        \fmf{fermion,label=$L$,label.side=right}{o1,v3}
        \fmf{fermion,label=$l_R$,label.side=right}{v2,o2}
        \fmf{fermion}{v2,v1}
        \fmf{fermion}{v3,v2}
        \fmffreeze
        \fmf{boson}{v1,v3}
    \end{fmfgraph*}
    \end{fmffile}
    \caption{Example of electroweak corrections to Type A contact
    interactions. All permutations are required including wavefunction renormalization;
    the vector bosons are the
    $W_\mu^{1,2,3}$ and $B_\mu$.
    \label{fig4}}
    \end{center}
\end{figure}
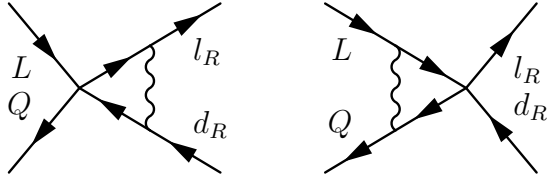

We start by considering $\textnormal{SU}(2) \times \textnormal{U}(1)$ invariant four-fermion
contact interactions that are generation independent and flavour diagonal (see figure \ref{fig3}
and figure \ref{fig6}). We will discuss the effects of generation dependence in
section \ref{lasts}.
We consider two types of scalar operators in order to
facilitate comparison with the direct experimental constraints. Type A ($O_A$) have left-handed neutrinos
in the final state while Type B ($O_B$) have right-handed (sterile) neutrinos.
These interactions appear as extensions to the standard model Lagrangian involving
non-renormalizable operators,
\be
\mathcal{L}_{\textnormal{scalar}} = \frac{s_A}{\Lambda^2} O_A + \frac{s_B}{\Lambda^2} O_B
\label{lag}
\ee
where $s_A$ and $s_B$ are undetermined scalar couplings.
From these interactions, electroweak radiative corrections (see figure \ref{fig4} and figure \ref{fig7}) can in
principle induce pseudoscalar interactions. We retain corrections up to order $\frac{1}{\Lambda^2}$ and from
this analysis we extract the anomalous dimension matrix.

\subsection{Type A Operator Analysis: $O_A$}
\label{suba}
The operators of Type A are as follows,
\bea
O_1 &=&[\bar e_R L][\bar Q d_R] \\
O_2 &=& [\bar e_R L][\bar u_R Q],
\eea
(where the SU(2) indices have been suppressed) such that the pure scalar interaction is
\be
O_A = O_1 + O_2.
\ee
%%%%%%%%%%%%%%%%%%%%%%%%%%%%%%%%%%%%%%%%%%
Since we are assuming that at the scale $\Lambda$ there is a pure scalar interaction, we take
$O_1$ and $O_2$ to enter the theory at the high scale with equal weight.

\begin{figure}[!h]
\newlength{\picw}
\setlength{\picw}{3in}
 \begin{center}
\subfigure[][]{\resizebox{\picw}{!}{\includegraphics{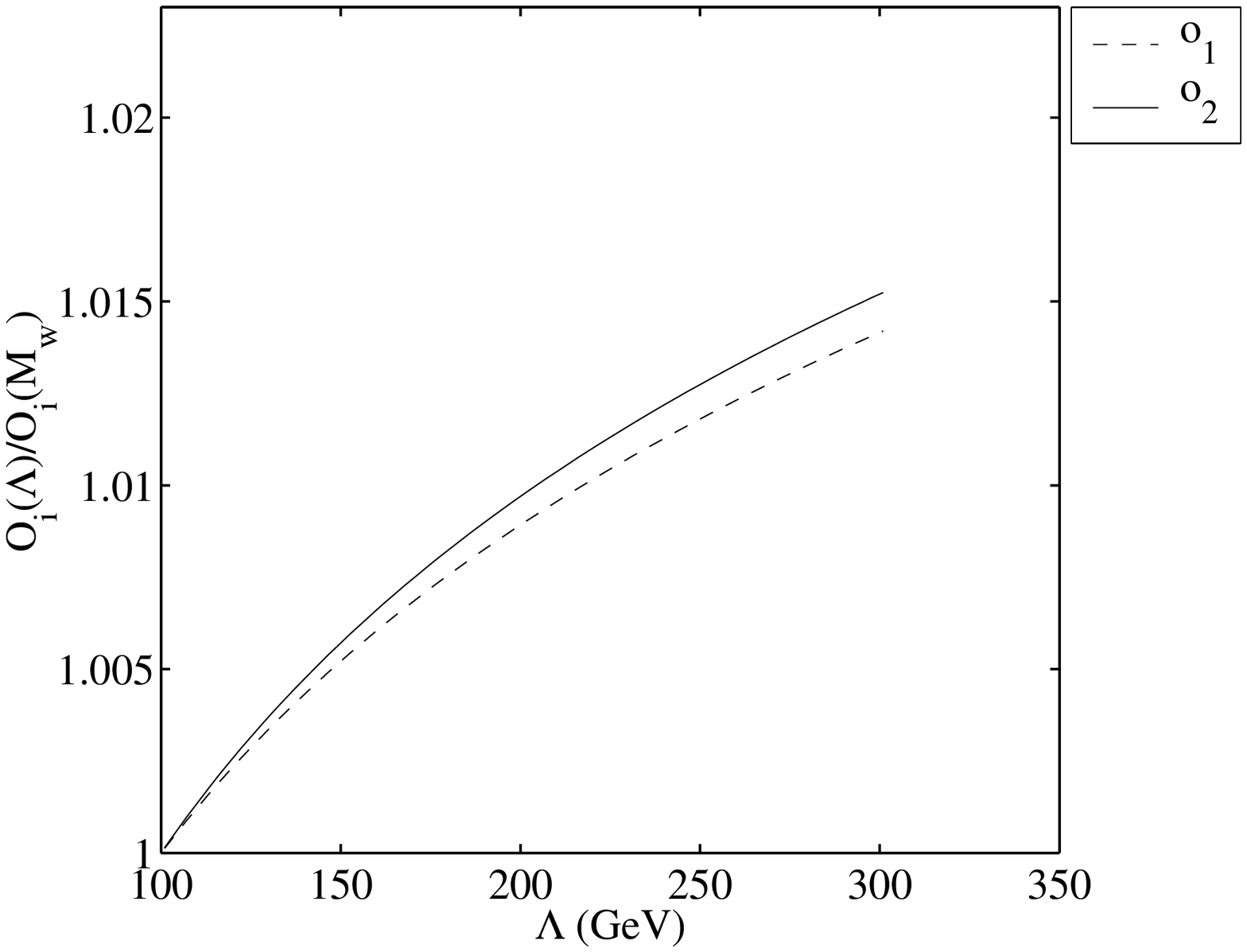}}}
 \subfigure[][]{\resizebox{\picw}{!}{\includegraphics{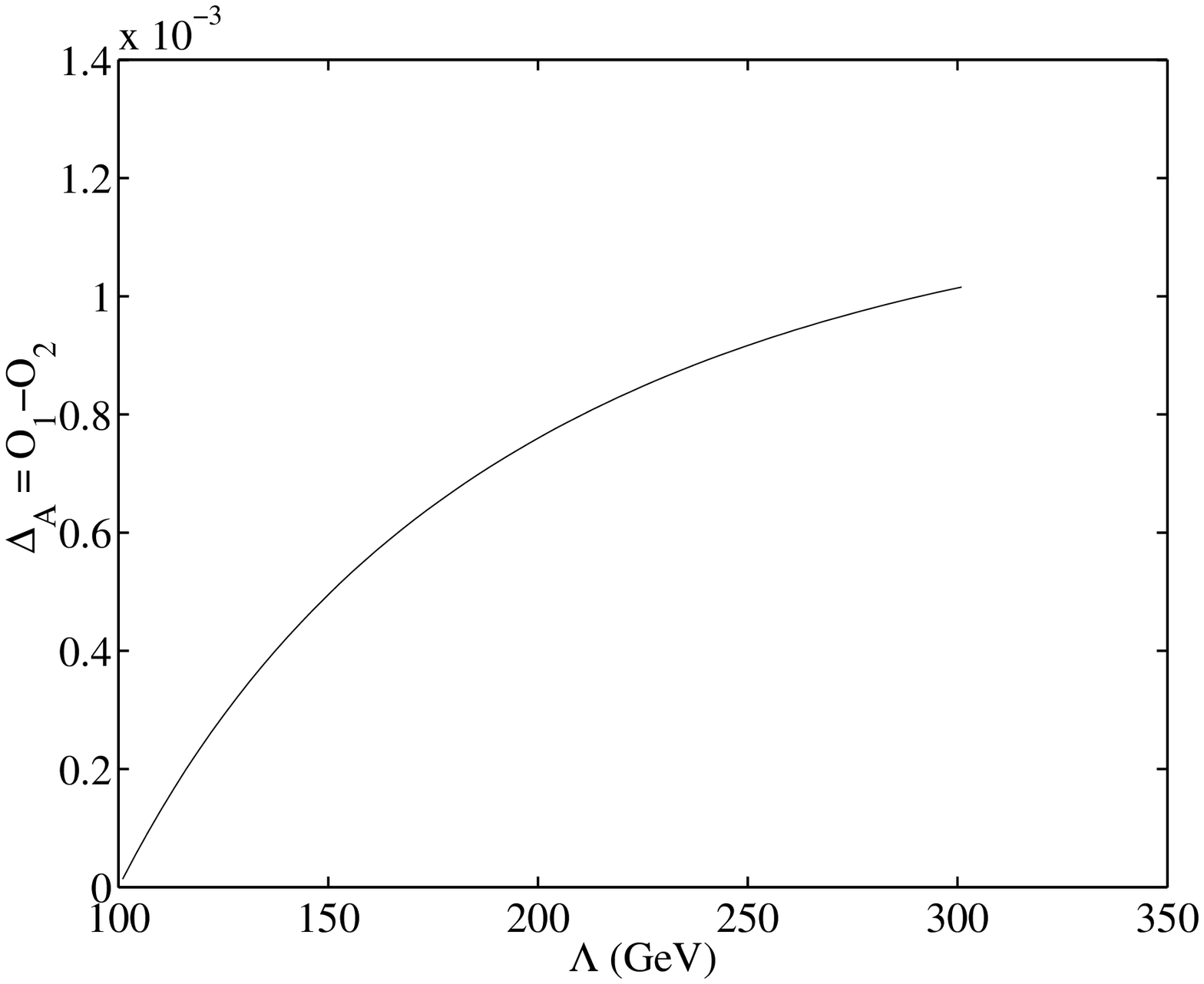}}}
 \end{center}
\caption{Type A operator RGE analysis. Panel (a) shows how each operator evolves with scale.
Panel (b) displays the induced pseudoscalar proportionality factor.}
\label{fig5}
\end{figure}

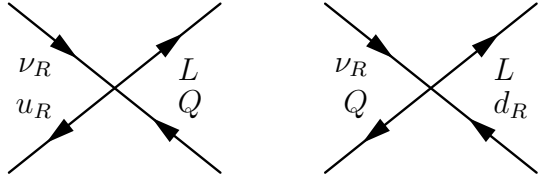
\begin{figure}
    \begin{center}
    \begin{fmffile}{corr5}
    \begin{fmfgraph*}(80,80)
        \fmfbottom{i1,i2}
        \fmftop{o1,o2}
        \fmf{fermion,label=$u_R$,label.side=right}{v1,i1}
        \fmf{fermion,label=$Q$,label.side=right}{i2,v1}
        \fmf{fermion,label=$\nu_R$,label.side=right}{o1,v1}
        \fmf{fermion,label=$L$,label.side=right}{v1,o2}
    \end{fmfgraph*}
    \end{fmffile}
\makebox(20,20){}
    \begin{fmffile}{corr6}
    \begin{fmfgraph*}(80,80)
        \fmfbottom{i1,i2}
        \fmftop{o1,o2}
        \fmf{fermion,label=$Q$,label.side=right}{v1,i1}
        \fmf{fermion,label=$d_R$,label.side=right}{i2,v1}
        \fmf{fermion,label=$\nu_R$,label.side=right}{o1,v1}
        \fmf{fermion,label=$L$,label.side=right}{v1,o2}
    \end{fmfgraph*}
    \end{fmffile}
    \caption{$\mathcal{O}_1$ and $\mathcal{O}_2$, Type B contact interactions
    \label{fig6}}
    \end{center}
\end{figure}

\begin{figure}
    \begin{center}
    \begin{fmffile}{corr7}
    \begin{fmfgraph*}(80,80)
        \fmfbottom{i1,i2}
        \fmftop{o1,o2}
        \fmf{fermion,label=$u_R$,label.side=right}{v1,i1}
        \fmf{fermion,label=$Q$,label.side=right}{i2,v2}
        \fmf{fermion,label=$\nu_R$,label.side=right}{o1,v1}
        \fmf{fermion}{v2,v1}
        \fmf{fermion}{v1,v3}
        \fmf{fermion,label=$L$,label.side=right}{v3,o2}
        \fmffreeze
        \fmf{boson}{v3,v2}
    \end{fmfgraph*}
    \end{fmffile}
\makebox(20,20){}
    \begin{fmffile}{corr8}
    \begin{fmfgraph*}(80,80)
        \fmfbottom{i1,i2}
        \fmftop{o1,o2}
        \fmf{fermion,label=$Q$,label.side=right}{v1,i1}
        \fmf{fermion,label=$d_R$,label.side=right}{i2,v2}
        \fmf{fermion,label=$\nu_R$,label.side=right}{o1,v1}
        \fmf{fermion,label=$L$,label.side=right}{v3,o2}
        \fmf{fermion}{v2,v1}
        \fmf{fermion}{v1,v3}
        \fmffreeze
        \fmf{boson}{v2,v3}
    \end{fmfgraph*}
    \end{fmffile}
    \caption{Example of electroweak corrections to Type B contact
    interactions. All permutations are required including wavefunction renormalization;
    the vector bosons are the
    $W_\mu^{1,2,3}$ and $B_\mu$.
    \label{fig7}}
    \end{center}
\end{figure}
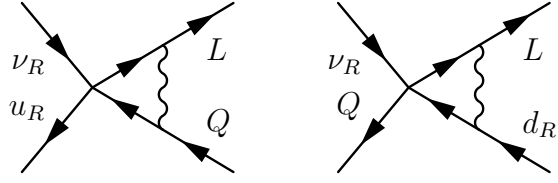

In calculating the anomalous dimension matrix a third operator is
generated through renormalization: the operator $O^\prime = [\bar
e_R Q][\bar u_R L]$ mixes with the other two. However, in order to
construct the matrix element for the pion decay amplitude, we need
to rotate the operators to a basis that has a definite matrix
element between the vacuum and the on-shell pion state. This
requires Fierz reordering,
\be
O^\prime= -\frac{1}{2}{O_2} +
(-\frac{1}{8})[\bar e_R \sigma_{\mu \nu} L][\bar u_R \sigma^{\mu
\nu}Q] \ee where we define \be O_3 \equiv (-\frac{1}{8})[\bar e_R
\sigma_{\mu \nu} L][\bar u_R \sigma^{\mu \nu}Q].
\ee
Note that $<0|O_3|\pi(p)> =0$. This leaves us with the following beta
functions,
\be
\mu \frac{\partial (\mathcal{O})}{\partial \mu} =
\frac{1}{32 \pi^2} \mathbf{\gamma} \mathcal{O}
\ee where,
\be
\mathcal{O} = \left(\begin{array}{c} O_1 \\ O_2 \\ O_3
\end{array} \right)
\ee and
\be
\mathbf{\gamma}= \left[ \begin{array}{ccc} 6 g^2 + \frac{98}{9} g^{\prime 2} & 0 & 0 \\
0 & 6g^2 + \frac{128}{9} g^{\prime 2} & 6g^2 + 10g^{\prime 2} \\ 0
& \frac{9}{2} g^2 + \frac{15}{2} g^{\prime 2} & 12 g^2 +
\frac{103}{9} g^{\prime 2} \end{array} \right].
\ee
The constants $g^\prime$ and $g$ are the $\textnormal{U}(1)$ and
$\textnormal{SU}(2)$ coupling constants, respectively. The results
of the numerical integration of the renormalization group
equations are displayed in figure \ref{fig5}. $O_1$ and $O_2$
start out with equal amplitude at the scale $\Lambda$. They are
then renormalized to the weak scale of roughly 100 GeV. In the
first panel the x-axis indicates the starting scale $\Lambda$,
i.e. the scale of new physics. The y-axis indicates the amount
each operator is suppressed in running from the scale $\Lambda$ to
the weak scale. Each operator renormalizes differently and the
splittings give rise to the pseudoscalar interaction. If the
scale $\Lambda$ is at or very near the weak scale then threshold
effects become important, which we will discuss in the following section.
The second panel plots the
difference of $O_1$ and $O_2$ as a function of scale. This
difference is proportional to the amount of pseudoscalar
interaction induced.

\subsection{Type B Operator Analysis: $O_B$}
\label{subb}
The Type B operators are as follows,
\bea
O_1 &=& [\bar L \nu_R] [\bar Q d_R] \\
O_2 &=& [\bar L \nu_R] [\bar u_R Q]
\eea
(where the SU(2) indices have been suppressed) with
\be
O_B=O_1+O_2.
\ee
%%%%%%%%%%%%%%%%%%%%%%%%%%%%%%%%%%%%%%%%%%
We assume that the interaction at the scale $\Lambda$ is purely
scalar as in the Type A scenario. Again operator mixing is present
with a third induced operator, namely $O^\prime =[\bar L d_R][\bar
Q \nu_R]$ which must be rotated as before into the appropriate
basis: $O^\prime = -\frac{1}{2} O_2 + (-\frac{1}{8})[\bar L
\sigma^{\mu \nu} \nu_R][\bar Q \sigma_{\mu \nu} d_R]$ where $O_3 =
(-\frac{1}{8})[\bar L \sigma^{\mu \nu} \nu_R][\bar Q \sigma_{\mu
\nu} d_R]$. We extract the following anomalous dimension matrix:
\be
\mu \frac{\partial (\mathcal{O})}{\partial \mu} = \frac{1}{32
\pi^2} \mathbf{\gamma} \mathcal{O} \ee where, \be \mathcal{O} =
\left(\begin{array}{c} O_1 \\ O_2 \\ O_3 \end{array} \right)
\ee
and
\be
\mathbf{\gamma}= \left[ \begin{array}{ccc} 6 g^2 + \frac{38}{9} g^{\prime 2} & 0 & 0 \\
0 & 6g^2 + \frac{11}{9} g^{\prime 2} & 6g^2 - \frac{2}{3}g^{\prime 2} \\ 0 & \frac{9}{2} g^2 - \frac{1}{3} g^{\prime
2} & 12 g^2 + \frac{34}{9} g^{\prime 2} \end{array} \right].
\ee
\begin{figure}[!ht]
\newlength{\pice}
\setlength{\pice}{3in}
 \begin{center}
\subfigure[][]{\resizebox{\pice}{!}{\includegraphics{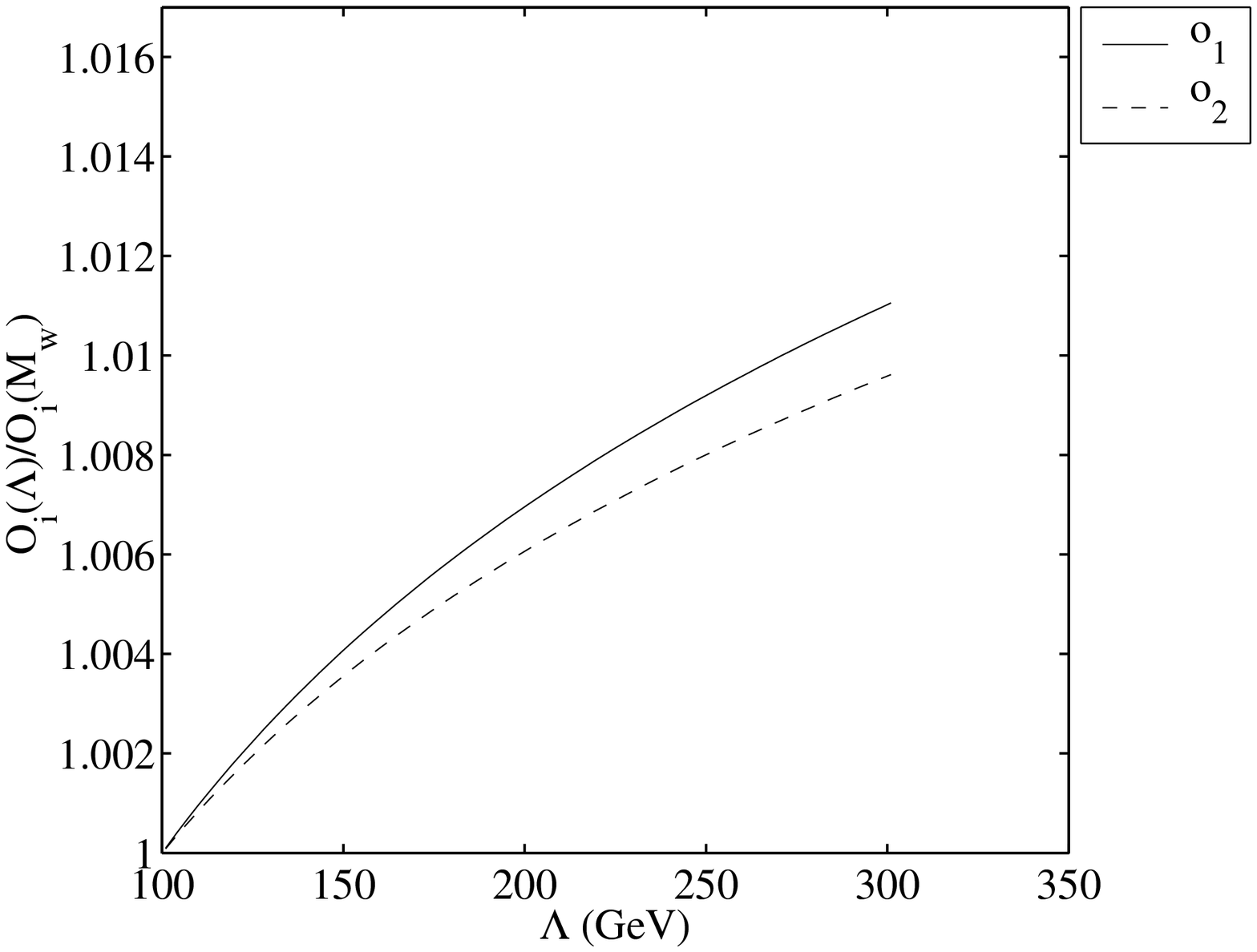}}}
\subfigure[][]{\resizebox{\pice}{!}{\includegraphics{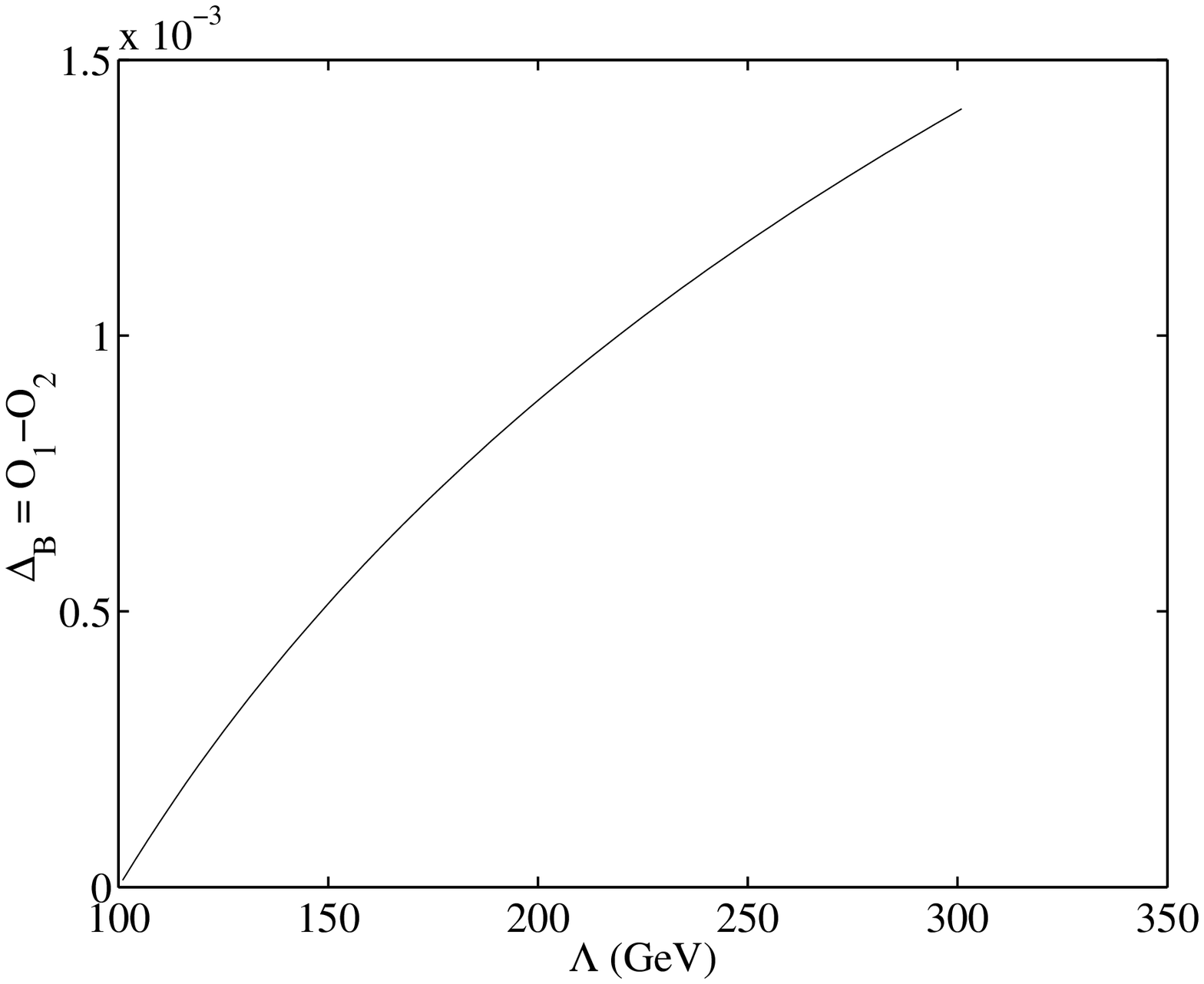}}}
 \end{center}
\caption{Type B operator RGE analysis. Panel (a) shows how each operator evolves with scale.
Panel (b) displays the induced pseudoscalar proportionality factor.}
\label{fig8}
\end{figure}
The results of the numerical integration of the renormalization group equations are displayed in figure \ref{fig8}.
As we have seen before in section \ref{suba} the graphs in figure \ref{fig8} illustrate the effects
of renormalization on the operators $O_1$ and $O_2$ when they enter with the same amplitude at the scale $\Lambda$.

In both Type A and B scalar interactions we see that renormalization effects induce a pseudoscalar
interaction. The size of the pseudoscalar interaction depends on how far the scale $\Lambda$ is
from the
weak scale. The larger the scale separation is, the larger the induced pseudoscalar proportionality factor becomes.
The effective pseudoscalar couplings, which we denoted as $\rho$ and $\rho^\prime$ in section \ref{pionp} , are given by,
\bea
\label{ps}
\rho &=& s_A \Delta_A(\Lambda) \nonumber \\
\rho^\prime &=& s_B \Delta_B (\Lambda)
\label{RHO}
\eea
where $\Delta_A$ and $\Delta_B$ are the renormalization group factors induced
from the running from the scale $\Lambda$ down to the weak scale ($\Delta_A$ and $\Delta_B$ are plotted
in the second panel of figure \ref{fig5} and figure \ref{fig8}). The factors $s_A$ and $s_B$ are the undetermined
scalar coupling constants introduced in eq.(\ref{lag}). Since the pseudoscalar is induced from a scalar interaction
we are now in a position to place limits on the magnitude of the scalar coupling from pion physics; 
the scalar couplings $s_A$ and $s_B$ at the scale of the new physics $\Lambda$
are now constrained by the requirement that $\rho$ and $\rho^\prime$ satisfy eq.(\ref{csteq}).

A comment on QCD corrections is in order. QCD is a parity invariant theory and
therefore QCD corrections cannot induce a pseudoscalar interaction by themselves. In our
analysis, the induced pseudoscalar arises from the difference of two
operators that initially combined to give a purely scalar interaction and the QCD
corrections will affect the two operators in the same way.
The QCD corrections can only adjust this difference by an overall multiplicative factor.
This is true for both operators of Type A and B.
However, in section \ref{betac} we compare the direct experimental constraints on scalar couplings from
$\beta$ decay to the indirect constraints on the renormalization induced pseudoscalar interactions
from pion decay. Since the same scalar operators are involved in both processes, the QCD effects are the same
for each case and therefore will cancel in a comparison of the relative strengths of the limits from the two processes.
The largest part of the QCD renormalization of the scalar operators (and hence of their weak interaction induced pseudoscalar
difference) will come from the QCD induced running from the weak scale down to 
the chiral symmetry breaking scale, of order $4\pi f_\pi \approx 1 \textnormal{GeV}$
\cite{Weinberg:1979}, where we take the pion decay matrix element using PCAC.
The correction to each of the operators can be computed through the QCD renormalization group running of these operators,
\bea
O_{A,B}(1 \hspace{.5mm} \textnormal{GeV}) &=& \left(\frac{\alpha_s(1 \hspace{.5mm} \mathrm{GeV}^2)}{\alpha_s(M_W^2)}\right)^{4/21} O_{A,B}(M_w) \nonumber \\
&\approx& 1.3 \hspace{1mm} O_{A,B}(M_w)
%O_{A,B}(M_w) &=& \left( \frac{\ln\left(\frac{M_W^2}{\Lambda_{QCD}^2}\right)}
%{\ln \left(\frac{1\hspace{.5mm}\mathrm{GeV}^2}{\Lambda
%_{QCD}^2} \right)}\right)^{4/21} \hspace{.5mm} O_{A,B} \nonumber \\
%&\approx& 1.3 \hspace{.5mm} O_{A,B}(1 \hspace{.5mm} \textnormal{GeV})
\eea
for $\Lambda_{QCD} = 200 \hspace{.5mm}\textnormal{MeV}$.
The induced pseudoscalar, which is proportional to $\Delta_{A,B}$, will be
enhanced by this factor of 1.3.
%\bea
%\Delta_{A,B} &\rightarrow& (1.3) \space \Delta_{A,B}.
%\label{QCD}
%\eea

\section{Pseudoscalar Interactions From Threshold Effects}
 \label{modelcal}

\begin{figure}
    \begin{center}
    \begin{fmffile}{box1}
    \begin{fmfgraph*}(80,150)
        \fmfbottom{i1,i2}
        \fmftop{o1,o2}
        \fmf{fermion,label=$u$,label.side=left}{v1,i1}
            \fmf{fermion,label=$d$,label.side=left}{i2,v2}
        \fmf{fermion,label=$\nu_l$,label.side=left}{o1,v3}
        \fmf{fermion,label=$l$,label.side=left}{v4,o2}
        \fmf{boson,label=$Z^0$,label.side=left}{v1,v3}
        \fmf{scalar,label=$S^-$,label.side=left}{v2,v4}
        \fmffreeze
        \fmf{fermion,label=$u$,label.side=left}{v2,v1}
        \fmf{fermion,label=$\nu_l$,label.side=left}{v3,v4}
    \end{fmfgraph*}
    \end{fmffile}
    %\end{center}
\makebox(20,20){}
    \begin{fmffile}{box2}
    \begin{fmfgraph*}(80,150)
        \fmfbottom{i1,i2}
        \fmftop{o1,o2}
        \fmf{fermion,label=$u$,label.side=left}{v1,i1}
            \fmf{fermion,label=$d$,label.side=left}{i2,v2}
        \fmf{fermion,label=$\nu_l$,label.side=left}{o1,v3}
        \fmf{fermion,label=$l$,label.side=left}{v4,o2}
        \fmf{scalar}{v1,v3}
        \fmf{boson}{v2,v4}
        \fmffreeze
        \fmf{fermion,label=$d$,label.side=left}{v2,v1}
        \fmf{fermion,label=$l$,label.side=left}{v3,v4}
    \end{fmfgraph*}
    \end{fmffile}
    %\end{center}
\makebox(20,20){}
    \begin{fmffile}{box3}
    \begin{fmfgraph*}(80,150)
        \fmfbottom{i1,i2}
        \fmftop{o1,o2}
        \fmf{fermion,label=$u$,label.side=left}{v1,i1}
            \fmf{fermion,label=$d$,label.side=left}{i2,v2}
        \fmf{fermion,label=$\nu_l$,label.side=left}{o1,v3}
        \fmf{fermion,label=$l$,label.side=left}{v4,o2}
        \fmf{phantom}{v1,v3}
        \fmf{phantom}{v2,v4}
        \fmffreeze
        \fmf{fermion,label=$d$,label.side=left}{v2,v1}
        \fmf{fermion,label=$\nu_l$,label.side=left}{v3,v4}
        \fmf{scalar,rubout}{v1,v4}
        \fmf{boson}{v2,v3}
    \end{fmfgraph*}
    \end{fmffile}
    %\end{center}
\makebox(20,20){}
    \begin{fmffile}{box4}
    \begin{fmfgraph*}(80,150)
        \fmfbottom{i1,i2}
        \fmftop{o1,o2}
        \fmf{fermion,label=$u$,label.side=left}{v1,i1}
            \fmf{fermion,label=$d$,label.side=left}{i2,v2}
        \fmf{fermion,label=$\nu_l$,label.side=left}{o1,v3}
        \fmf{fermion,label=$l$,label.side=left}{v4,o2}
        \fmf{phantom}{v1,v3}
        \fmf{phantom}{v2,v4}
        \fmffreeze
        \fmf{fermion,label=$u$,label.side=left}{v2,v1}
        \fmf{fermion,label=$l$,label.side=left}{v3,v4}
        \fmf{scalar,rubout}{v2,v3}
        \fmf{boson}{v1,v4}
    \end{fmfgraph*}
    \end{fmffile}
    \end{center}
    \caption{Dressed $Z^0$ exchange diagrams.}
    \label{fig1}
\end{figure}
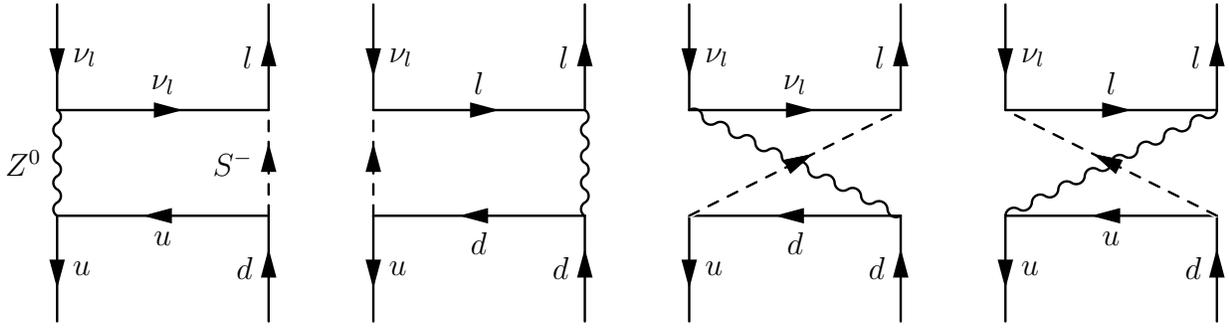

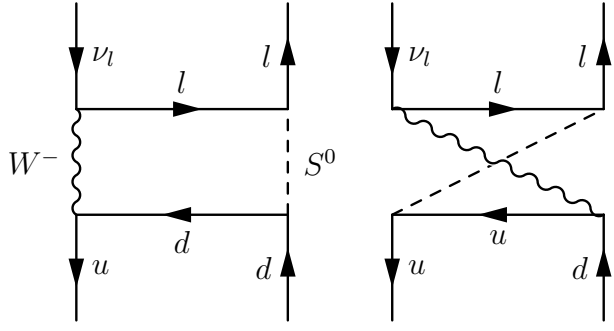
\begin{figure}[ht]
    \begin{center}
    \begin{fmffile}{box10}
    \begin{fmfgraph*}(80,150)
        \fmfbottom{i1,i2}
        \fmftop{o1,o2}
        \fmf{fermion,label=$u$,label.side=left}{v1,i1}
            \fmf{fermion,label=$d$,label.side=left}{i2,v2}
        \fmf{fermion,label=$\nu_l$,label.side=left}{o1,v3}
        \fmf{fermion,label=$l$,label.side=left}{v4,o2}
        \fmf{boson,label=$W^-$,label.side=left}{v1,v3}
        \fmf{dashes,label=$S^0$,label.side=left }{v2,v4}
        \fmffreeze
        \fmf{fermion,label=$d$,label.side=left}{v2,v1}
        \fmf{fermion,label=$l$,label.side=left}{v3,v4}
    \end{fmfgraph*}
    \end{fmffile}
    %\end{center}
\makebox(20,20){}
    \begin{fmffile}{box12}
    \begin{fmfgraph*}(80,150)
        \fmfbottom{i1,i2}
        \fmftop{o1,o2}
        \fmf{fermion,label=$u$,label.side=left}{v1,i1}
            \fmf{fermion,label=$d$,label.side=left}{i2,v2}
        \fmf{fermion,label=$\nu_l$,label.side=left}{o1,v3}
        \fmf{fermion,label=$l$,label.side=left}{v4,o2}
        \fmf{phantom}{v1,v3}
        \fmf{phantom}{v2,v4}
        \fmffreeze
        \fmf{fermion,label=$u$,label.side=left}{v2,v1}
        \fmf{fermion,label=$l$,label.side=left}{v3,v4}
        \fmf{boson,rubout}{v2,v3}
        \fmf{dashes}{v1,v4}
    \end{fmfgraph*}
    \end{fmffile}
    \end{center}
    \caption{Dressed $W$ exchange diagrams.}
    \label{fig2}
\end{figure}

\begin{figure}[!h]
    \begin{center}
    \begin{fmffile}{snout1}
    \begin{fmfgraph*}(100,100)
        \fmfbottom{i1,i2}
        \fmftop{o1,o2}
        \fmf{fermion,label=$u$,label.side=left}{v1,i1}
        \fmf{fermion,label=$d$,label.side=left}{i2,v2}
        \fmf{fermion}{v3,v1}
        \fmf{fermion}{v2,v3}
        \fmf{dashes}{v3,v4}
        \fmf{fermion,label=$\nu_l$,label.side=right}{o1,v4}
        \fmf{fermion,label=$l$,label.side=right}{v4,o2}
        \fmffreeze
        \fmf{photon}{v2,v1}
    \end{fmfgraph*}
    \end{fmffile}
    \makebox(20,20){}
%%%%%%%%%%%%%%%%%%%%%%%%%%%%%%%%%%%%%%%%%%%%%%
    \begin{fmffile}{snout2}
    \begin{fmfgraph*}(100,100)
        \fmfbottom{i1,i2}
        \fmftop{o1,o2}
        \fmf{plain,label=$d$,label.side=right}{v4,i2}
        \fmf{fermion}{v4,v2}
        \fmf{fermion,label=$u$,label.side=left}{v1,i1}
        \fmf{dashes}{v2,v3}
        \fmf{dashes}{v3,v5}
        \fmf{fermion}{v2,v1}
        \fmf{fermion,label=$\nu_l$,label.side=right}{o1,v5}
        \fmf{fermion,label=$l$,label.side=right}{v5,o2}
        \fmffreeze
        \fmf{photon}{v3,v1}
    \end{fmfgraph*}
    \end{fmffile}
    \makebox(20,20){}
%%%%%%%%%%%%%%%%%%%%%%%%%%%%%%%%%%%%%
    \begin{fmffile}{snout3}
    \begin{fmfgraph*}(100,100)
        \fmfbottom{i1,i2}
        \fmftop{o1,o2}
        \fmf{plain,label=$u$,label.side=left}{v4,i1}
        \fmf{fermion}{v2,v4}
        \fmf{fermion}{v1,v2}
        \fmf{fermion,label=$d$,label.side=left}{i2,v1}
        \fmf{dashes}{v2,v3}
        \fmf{dashes}{v3,v5}
        \fmf{fermion,label=$\nu_l$,label.side=right}{o1,v5}
        \fmf{fermion,label=$l$,label.side=right}{v5,o2}
        \fmffreeze
        \fmf{photon}{v3,v1}
    \end{fmfgraph*}
    \end{fmffile}
    \makebox(20,20){}
    \caption{Radiative corrections to the quark-scalar vertex. \label{divs}}
        \end{center}
\end{figure}
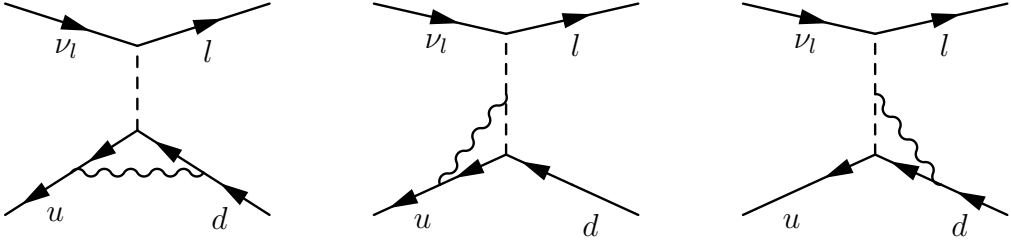

A limitation of the renormalization group operator analysis
of the last section is its inapplicability if the scale of new physics is at or very near
the electroweak scale. In this case, threshold effects become the dominate contribution.
To estimate the threshold effects, we consider a toy model
where a VEVless scalar doublet is added to the standard model. Indeed it is
only for the exchange of a scalar doublet that we need to consider a
possible scale for new physics near the electroweak scale. For leptoquarks,
compositeness, and extra dimensional gravity, direct
experimental constraints imply \cite{PDG} that the scale $\Lambda$ of new
physics is sufficiently above the electroweak scale that RGE
running dominates threshold effects.
In principle, the addition of a VEVless scalar doublet can lead to both scalar and
pseudoscalar interactions in the tree level Lagrangian.
Since pseudoscalar interactions are directly constrained by tree level contributions
to pion decay and we are presently interested in limits on pure scalar interactions,
we arrange the couplings such that only scalar interactions arise at the scale of new physics,
\be
\mathcal{L} = (\lambda) \bar L e_R S + (\lambda^\prime) \bar Q d_R S - (\lambda^\prime) \bar  Q u_R \tilde S + \textnormal{h.c.}
\ee
where, $\lambda$ and $\lambda^\prime$ are the scalar couplings to the quarks and leptons respectively,
and $\tilde S = i \sigma^2 S$.
In this working example, the scalar interactions have the property that they couple in a universal and flavour diagonal manner with
undetermined scalar couplings to quarks and leptons. It is the charged scalar couplings that the $\beta$-decay experiments
constrain directly. The pseudoscalar interaction can potentially be induced at one loop through three classes of diagrams:
scalar-dressed Z exchange box diagrams, scalar-dressed W exchange box diagrams and radiative corrections to the quark vertex (see
figure \ref{fig1}, figure \ref{fig2} and figure \ref{divs}). The weak interactions do not respect parity
and the scalar interactions change chirality, thus diagrams of this form can potentially induce a
pseudoscalar interaction. To estimate the effect of the scalar on the branching ratio, we will make the
approximation that the quarks are massless and ignore external momenta.
Box diagrams that involve the Higgs or the Goldstone modes can be ignored since the couplings are
mass proportional and hence their contribution is small.

By explicit calculation we can show that while both the dressed W and Z exchange box diagrams give non-zero amplitudes,
their tensor structure is such that after taking the matrix element between the pion and the vacuum they give vanishing contributions.
In the vertex correction class of diagrams we are dealing with primitively divergent graphs (see figure \ref{divs}).
In order to obtain a conservative estimate of the induced pseudoscalar arising already from threshold effects, we can regulate 
the loop diagrams by cutting off the
loop momentum at the weak scale and integrate from $0$ to $M_Z$. 
Cutting off the loop momentum at $M_Z$ represents a conservative estimate,
in that the scale of new physics is at the weak scale and
therefore there is no scale separation for renormalization group
running proper. In this case we find a non-vanishing contribution.
The three graphs in figure \ref{divs} give the following result for the pion decay matrix element,
\bea
\mathcal{M}_{\mathrm{Vertex}} &=& -\frac{\sqrt{2} g^2 \tilde f_\pi \lambda \lambda^\prime}{64 \pi^2
\cos^2 \theta_w M_Z^2} \left[ \left(-\frac{4}{3} \sin^2 \theta_w \right) \ln(2) +
\cos(2 \theta_w) \left(\ln(2) - \frac{1}{2}\right)\right]
[\bar l(1-\gamma_5)\nu_l] \nonumber \\
&\approx& 0.13\frac{\sqrt{2} g^2 \tilde f_\pi \lambda \lambda^\prime}{64 \pi^2
\cos^2 \theta_w M_Z^2} 
[\bar l(1-\gamma_5)\nu_l].
\label{thres}
\eea

To get a second, independent, estimate of the threshold corrections, in a different renormalization prescription, we will imagine
integrating out the weak scale degrees of freedom (W, Z and scalars) to get an effective low-energy theory.
The resulting theory will have only dimension six four-fermion operators; to simplify our calculation let us imagine setting the
scalar masses just below the mass of the W and Z and integrating out the W and Z first and then immediately integrating out
the scalars, thus inducing the four-fermion operators. If we use a dimensionless regulator, the effective fermion-scalar theory after
integrating out the W and Z will have Yukawa couplings shifted by threshold effects neccessary to reproduce the residual effects of the 
W and Z in the resulting effective theory in which they are absent. 
These threshold corrections have been computed in \cite{Wright1, Wright2}. We then immediately integrate out the scalars, with
their corrected Yukawa couplings, to get the final low-energy effective theory of fermions with four-fermion couplings. 
Using the results for the threshold corrections for Yukawa couplings from \cite{Wright1, Wright2}, with
the gauge charge representations of our particles, and then immediately integrating out the scalars at the weak scale (which we take to be $M_Z$)
we get an effective induced interaction from the vertex corrections of:
\bea
\mathcal{M}_{\mathrm{Vertex}} &\approx& 0.08\frac{\sqrt{2} g^2 \tilde f_\pi \lambda \lambda^\prime}{64 \pi^2
\cos^2 \theta_w M_Z^2} 
[\bar l(1-\gamma_5)\nu_l]. \nonumber \\
&&
\label{thres1}
\eea
That the estimates of eq.(\ref{thres}) and eq.(\ref{thres1}), which use two entirely different 
regularization and renormalization prescriptions, agree to within a factor of two gives us confidence that estimates of the 
threshold corrections are of this order and are not artifacts of the regulator chosen. To be conservative,
we will use the estimate of eq.(\ref{thres1}) which in conjunction with 
eq.(\ref{csteq}) and in the absence of right-handed neutrinos gives,
\be
 -3 \times 10^{-2} \leq \frac{K_s}{G_F} \leq  6 \times 10^{-3}
\label{thresres}
\ee
where,
\be
|K_s| \equiv \frac{\lambda \lambda^\prime}{M_Z^2}.
\ee
The above calculation gives a conservative estimate of the amplitude, including only contributions from
threshold effects. We see in this toy example that even from
threshold effects alone a pseudoscalar interaction will be radiatively induced.

\section{Comparison with $\beta$-Decay Constraints}
\label{betac}
We can compare our bounds on scalar currents, with those arising in nuclear $\beta$-decay.
The effective Hamiltonian for allowed $\beta$-decay has the general Lorentz form
\cite{Jack:1957},
\bea
H&=& \frac{G_F}{\sqrt{2}}\left\{\right.(\bar\psi_p \gamma_\mu \psi_n)(C_V\bar\psi_e \gamma_\mu\psi_\nu +
C_V^\prime\bar\psi_e \gamma_\mu\gamma_5\psi_\nu)
\nonumber \\
&&+ (\bar\psi_p\gamma_\mu\gamma_5\psi_n)(C_A\bar\psi_e\gamma_\mu\psi_\nu +
C_A^\prime\bar\psi_e\gamma_\mu\gamma_5\psi_\nu) \nonumber \\
&&+(\psi_p\psi_n)(C_S\bar\psi_e\psi_\nu + C_S^\prime\bar\psi_e\gamma_5\psi_\nu)
\nonumber \\
&&+\frac{1}{2}(\bar\psi_p\sigma_{\lambda\mu}\psi_n)
(C_T\bar\psi_e\sigma_{\lambda\mu}\psi_\nu+C^\prime_T\bar\psi_e\sigma_{\lambda\mu}
\gamma_5\psi_\nu)\left.\right\}.
\label{Jack}
\eea
A pseudoscalar term has not been included since it vanishes to leading order in nuclear $\beta$ becay.
In the absence of right-handed currents, $C_i = C^\prime_i$ and
as we have mentioned before, we consider purely scalar interactions. (Note that in
the above, $\frac{1 + \gamma_5}{2}$ is taken to be the left projector. This is opposite
to our convention in the preceding sections. However by using this convention in
this section, it will be easier to compare with the $\beta$-decay literature.)
The transition probability per unit time is given by \cite{Jack:1957},
\be
w_{if} = \frac{\xi}{4\pi^3}{p_e E_e}(E_{max} -E_e) \left(1 + a
v_e \cos \theta + b \frac{2m_e}{E_e}\right) \sin \theta \hspace{1mm} d \theta
\ee
where $E_{max}$ is the maximum energy of the electron in beta decay,
$v_e=p_e/E_e$ and,
\bea
\xi &=& \frac{1}{2}|M_F|^2\left(|C_V|^2+|C_V^\prime|^2+|C_S|^2+|C_S^\prime|^2\right)
+\frac{1}{2}|M_{GT}|^2
\left(|C_A|^2+|C_A^\prime|^2+|C_T|^2+|C_T^\prime|^2\right) \nonumber \\
a\xi &=& \frac{1}{2}|M_F|^2\left(|C_V|^2+|C_V^\prime|^2-|C_S|^2-|C_S^\prime|^2\right)
-\frac{1}{6}
|M_{GT}|^2\left(|C_A|^2+|C_A^\prime|^2-|C_T|^2-|C_T^\prime|^2\right) \nonumber \\
b\xi &=& \frac{1}{2} \textnormal{Re} \left(C_S C_V^* + C_S^\prime C_V^{\prime
*}\right) |M_F|^2 + \frac{1}{2} \textnormal{Re}\left(C_T C_A^* + C_T^\prime
C_A^{\prime *} \right) |M_{GT}|^2.
\label{monster}
\eea
The angle, $\theta$, is the angle between the electron and neutrino momenta and $b$ is the
Fierz interference term.
The direct searches \cite{Garcia:1999da, Adelberger:1999ud, Adelberger:1993wq}
for scalar interactions in $\beta$-decay consider pure Fermi transitions $0^+ \rightarrow 0^+$
as the parameter $a$ has a particulary simple form. In this case the Gamow-Teller matrix elements are absent
and the Fermi matrix elements divide out,
\be
a=\frac{|C_V|^2 +|C_V^\prime|^2-|C_S|^2-|C_S^\prime|^2}
{|C_V|^2 +|C_V^\prime|^2+|C_S|^2+|C_S^\prime|^2}.
\ee
Since in the standard model $C_V=C_V^\prime=1$,
$a \neq 1$ implies evidence for an effective scalar
interaction.

We need to rewrite our expressions for scalar
interactions in terms of $\tilde C_s$ and
$\tilde C_s^\prime$ where $\tilde C_i = C_i/C_V$. The scalar couplings can be re-expressed,
\bea
S_A &=&  \frac{\Lambda^2 G_F \cos \theta_c}{\sqrt{2}} (\tilde C_s + \tilde C_s^\prime) \\
S_B &=&  \frac{\Lambda^2 G_F \cos \theta_c}{\sqrt{2}} (\tilde C_s
- \tilde C_s^\prime)
\eea
where the $S_A, S_B$ denote scalar interactions at the nucleon level. The operator
analysis of section \ref{opa} was completed with quarks, thus we need to
include the scalar form factor $<p|\bar u d|n>$ which can be estimated
from lattice calculations \cite{Liu:1998um},
$<p|\bar u d|n> \approx  0.65 \pm 0.09$. By saturating the error in this quantity,
we can obtain a conservative 2 $\sigma$ constraint equation on the scalar couplings from pion decay
(see eq.(\ref{csteq})),
\be
\label{mstc}
-1.0 \times 10^{-2} \leq \frac{1}{0.74}\frac{\tilde f_\pi \Delta_A}{f_\pi m_e}
\textnormal{Re}(\tilde C_s + \tilde C_s^\prime) +
\frac{1}{0.74^2}\frac{\Delta_A^2\tilde f_\pi^2}{f_\pi^2 m_e^2} |\tilde C_s +
\tilde C_s^\prime|^2 + \frac{1}{0.74^2}\frac{\Delta_B^2 \tilde f_\pi^2}{f_\pi^2
m_e^2} |\tilde C_s - \tilde C_s^\prime|^2  \leq 2.2 \times
10^{-3}
\ee
If we include only left-handed neutrinos in the theory, we are constrained to lie along the line $\tilde C_s=\tilde C_s^\prime$ whereas if we include only
right-handed neutrinos we are forced to lie along $\tilde
C_s=-\tilde C_s^\prime$. We can now examine a few special cases.

\begin{figure}[ht!]
\newlength{\picwidth}
\setlength{\picwidth}{3in}
 \begin{center}
\subfigure[][]{\resizebox{\picwidth}{!}{\includegraphics{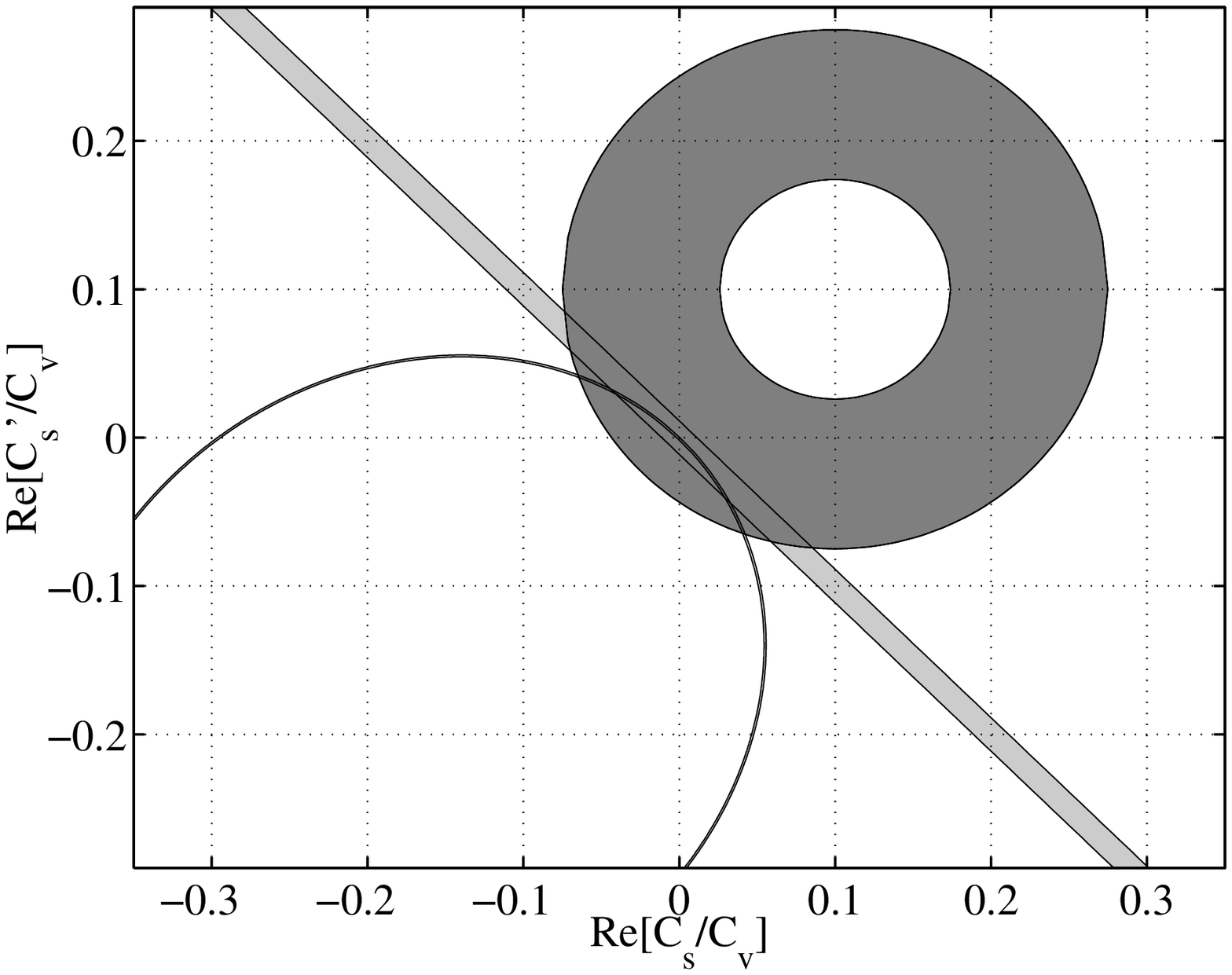}}}
 \subfigure[][]{\resizebox{\picwidth}{!}{\includegraphics{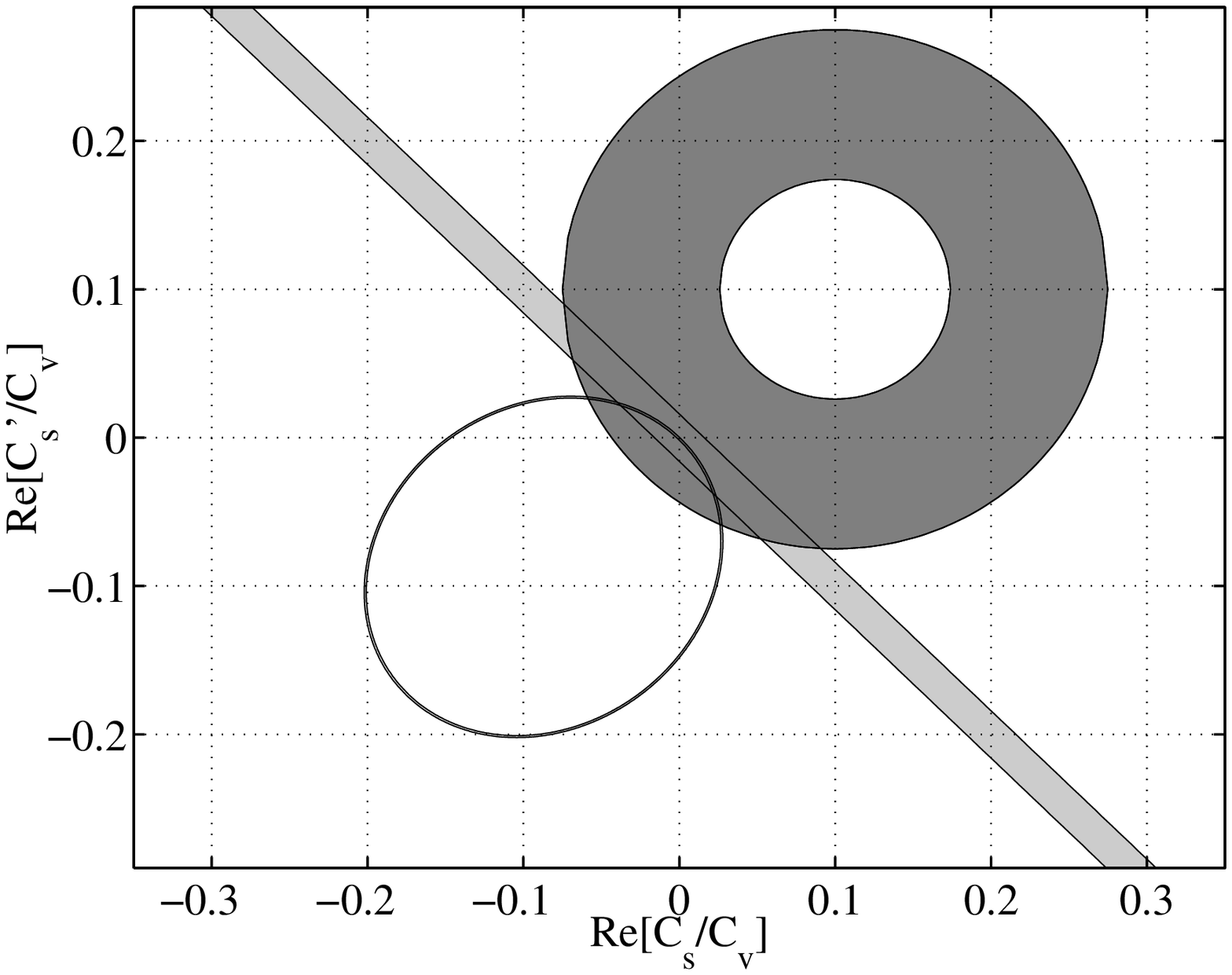}}}
 \subfigure[][]{\resizebox{\picwidth}{!}{\includegraphics{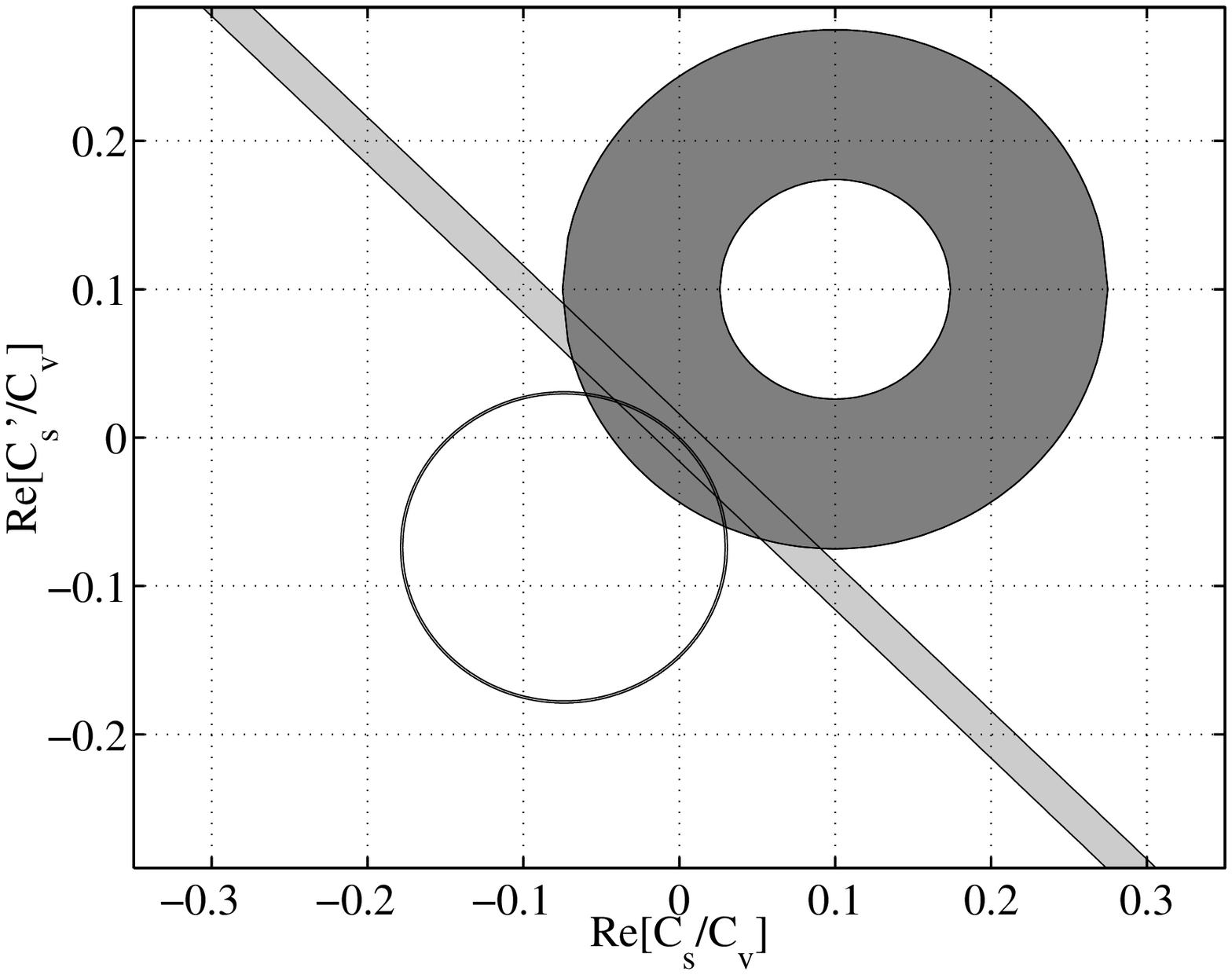}}}
 \end{center}
\caption{Constraint plots on the real parts of
$\tilde C_s$ and $\tilde C_s^\prime$ at $\Lambda=$200 GeV. Panel (a)
corresponds to a phase of $0^\circ$; panel (b) to $\pm 45^\circ$; and panel (c) to $45^\circ$ and $-45^\circ$
for $\tilde C_s$ and $\tilde C_s^\prime$ respectively.
The diagonal band is the experimental limit set by the b-Fierz
interference term from $\beta$-decay at the $90$\% confidence level and the solid annulus is the approximate
experimental bound given in \cite{Adelberger:1999ud}. In all cases,
the allowed region is the band between the two ellipses.
An enlargement of the figures is displayed in figure \ref{fig10}.}
\label{bigr}
\end{figure}

\begin{figure}[ht!]
\newlength{\picwidthm}
\setlength{\picwidthm}{3in}
 \begin{center}
\subfigure[][]{\resizebox{\picwidthm}{!}{\includegraphics{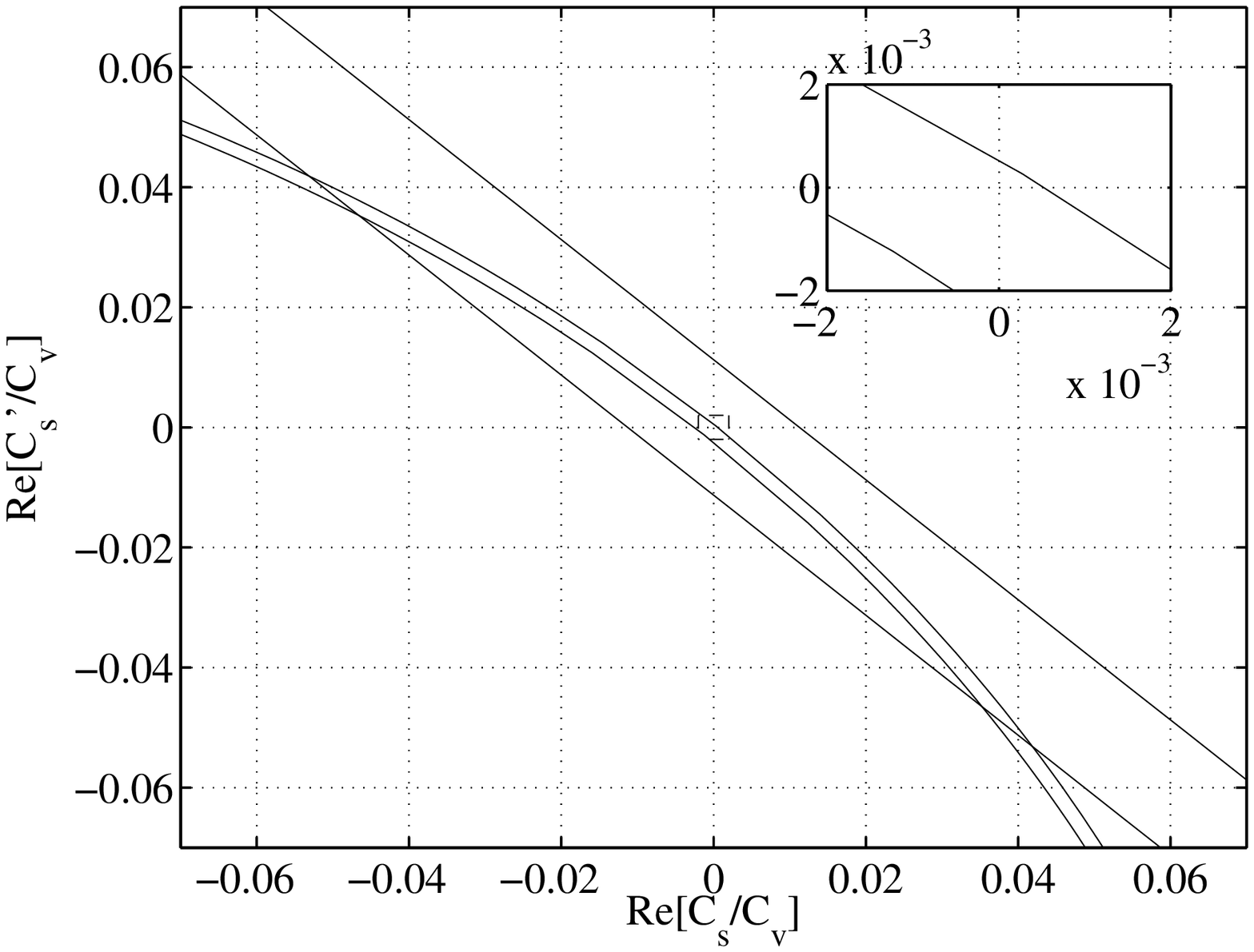}}}
 \subfigure[][]{\resizebox{\picwidthm}{!}{\includegraphics{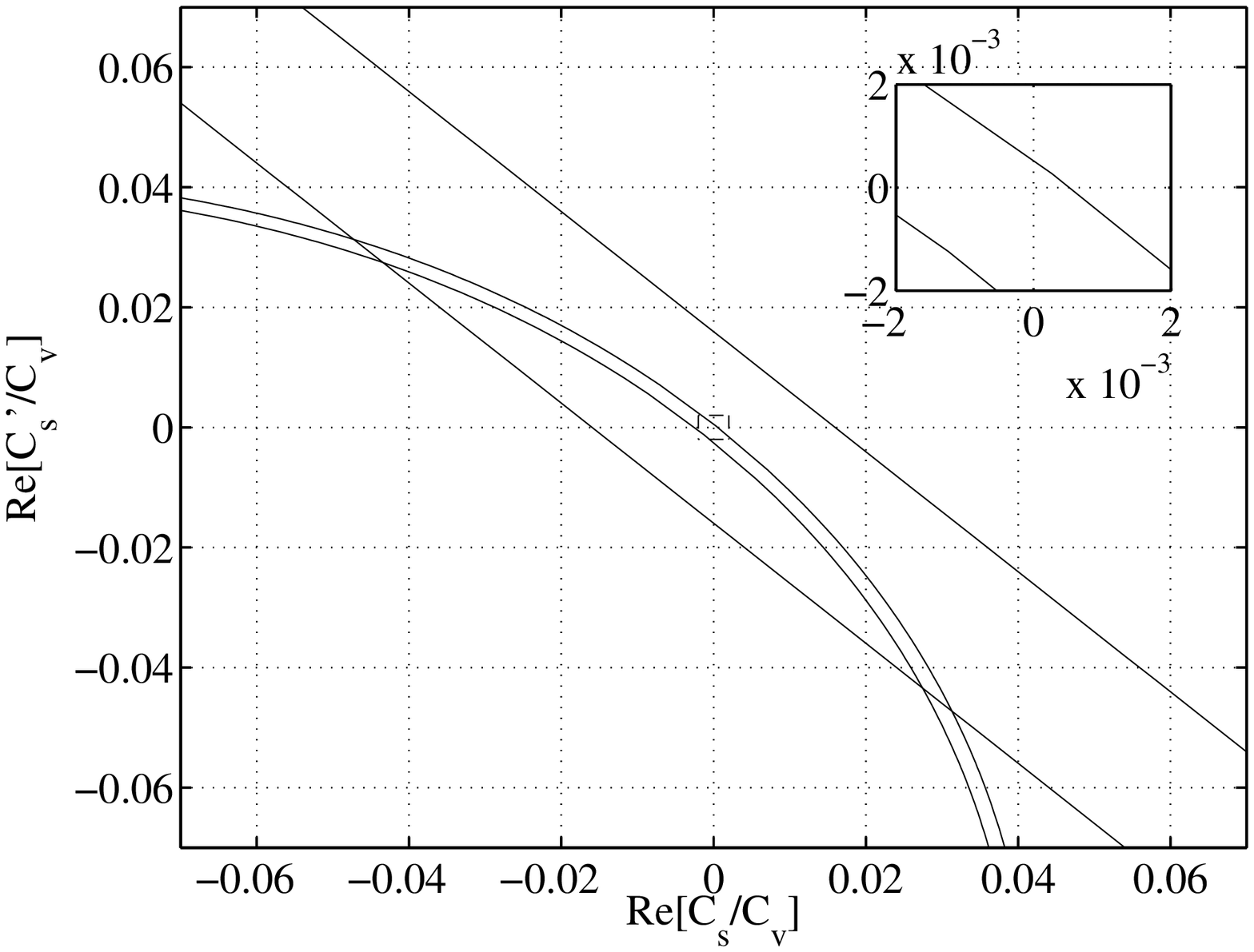}}}
 \subfigure[][]{\resizebox{\picwidthm}{!}{\includegraphics{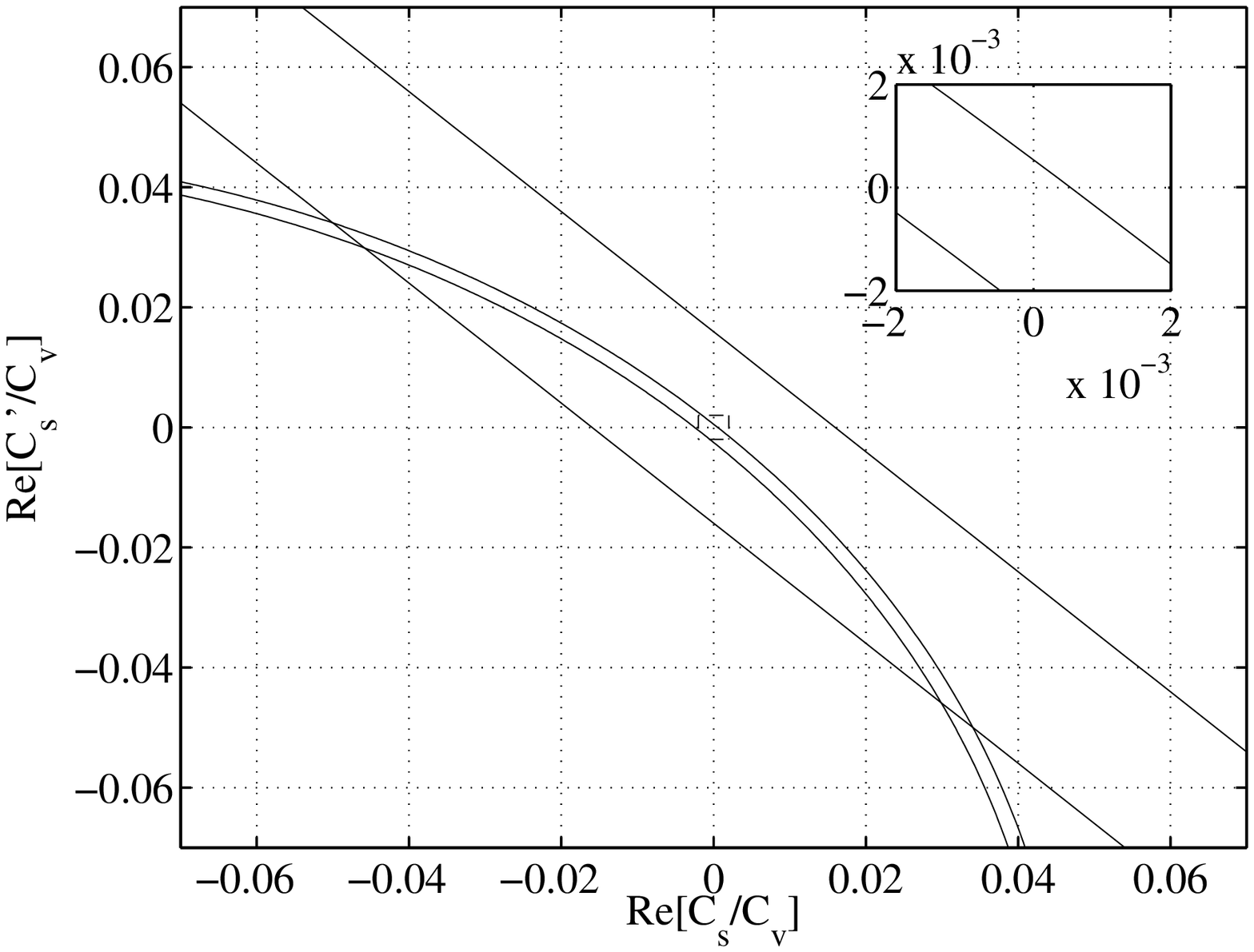}}}
 \end{center}
\caption{Constraint plots on the real parts of
$\tilde C_s$ and $\tilde C_s^\prime$ at $\Lambda=$200 GeV. Panel (a)
corresponds to a phase of $0^\circ$; panel (b) to $\pm 45^\circ$; and panel (c) to $45^\circ$ and $-45^\circ$
for $\tilde C_s$ and $\tilde C_s^\prime$ respectively.
The diagonal band is the experimental limit set by the b-Fierz
interference term from $\beta$-decay at the $90$\% confidence level. In all cases,
the allowed region is the band between the two ellipses. The enlarged area more clearly
shows the width of the region.}
\label{fig10}
\end{figure}

\begin{figure}[ht!]
\newlength{\picwidthn}
\setlength{\picwidthn}{3in}
 \begin{center}
 \subfigure[][]{\resizebox{\picwidthn}{!}{\includegraphics{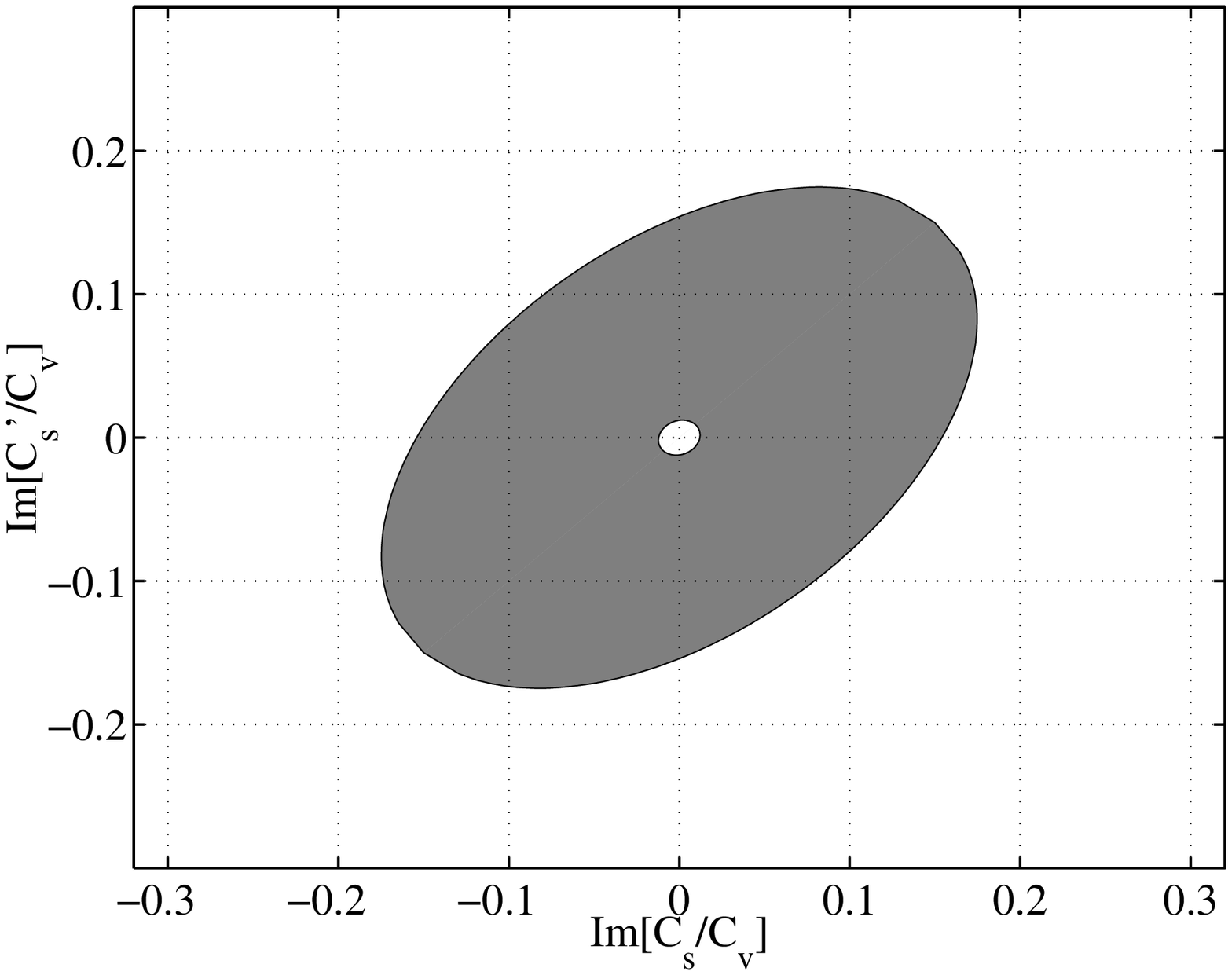}}}
 \subfigure[][]{\resizebox{\picwidthn}{!}{\includegraphics{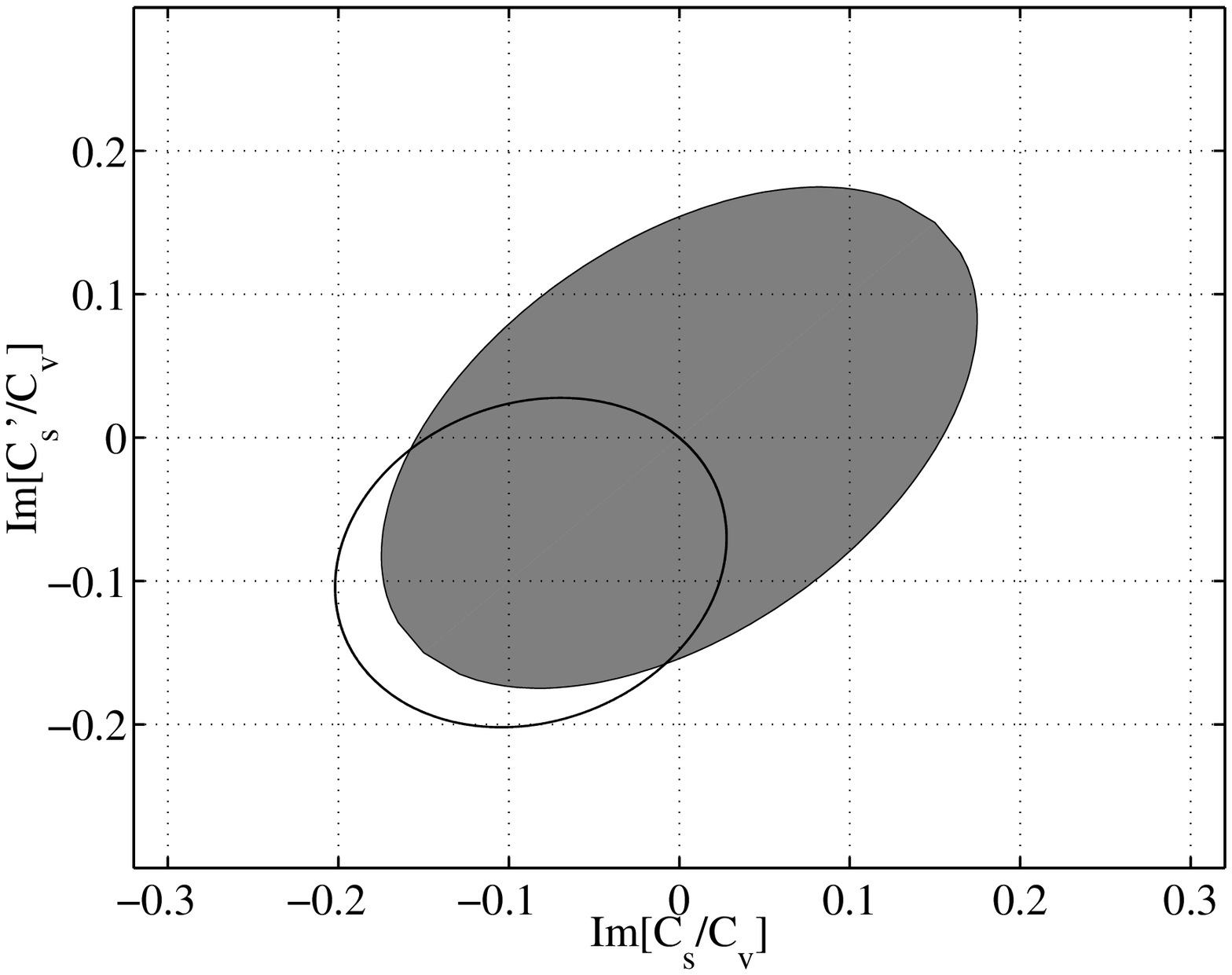}}}
 \subfigure[][]{\resizebox{\picwidthn}{!}{\includegraphics{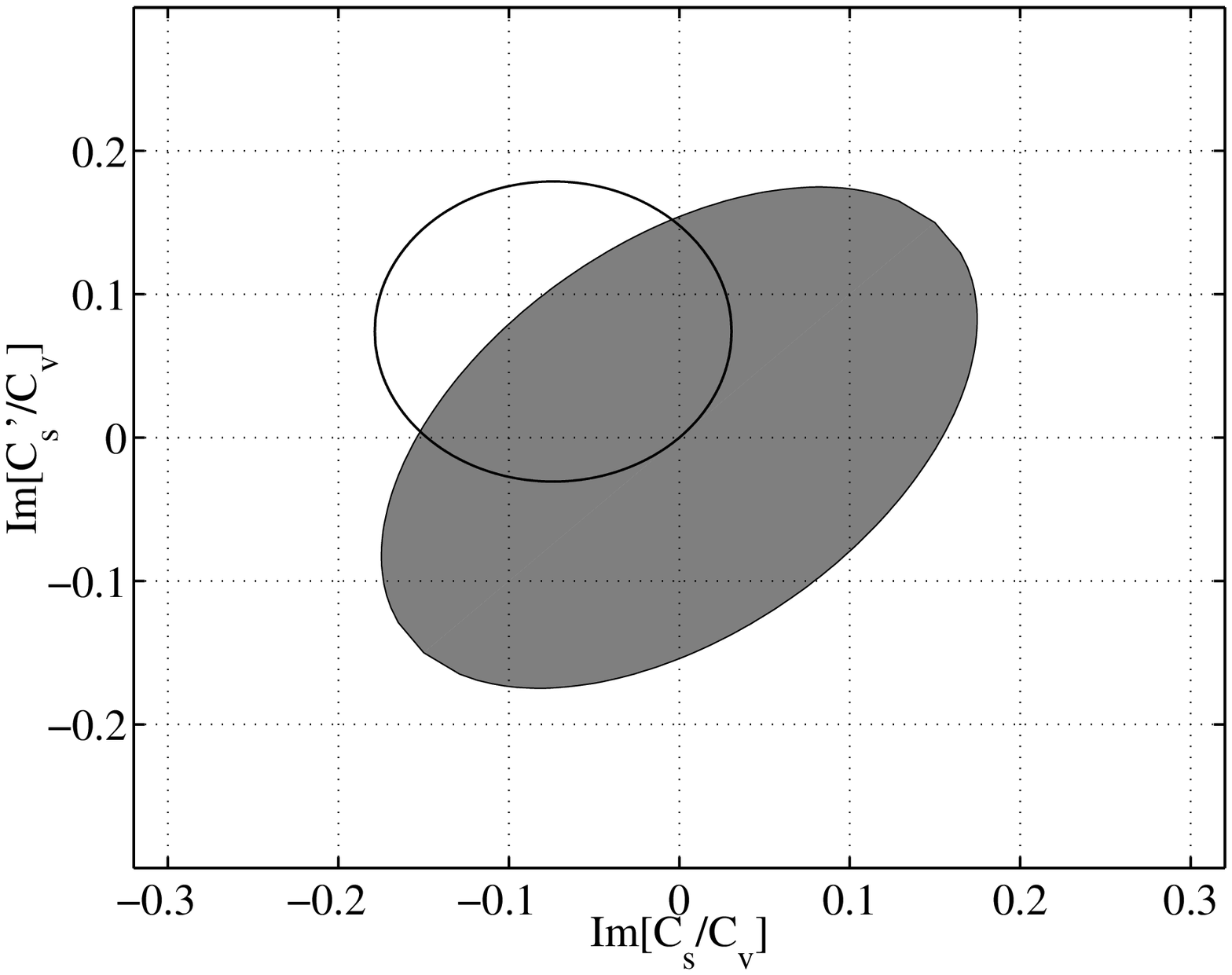}}}
 \end{center}
\caption{Constraint plots on the imaginary parts of
$\tilde C_s$ and $\tilde C_s^\prime$ at $\Lambda=$200 GeV. Panel (a)
corresponds to a phase of $\pm 90^\circ$; panel (b) to $\pm 45^\circ$;
and panel (c) to $45^\circ$ and $-45^\circ$ for $\tilde C_s$ and $\tilde C_s^\prime$ respectively.
The solid ellipse is the approximate experimental bound on the imaginary part of the couplings
assuming nothing about the phase \cite{Adelberger:1999ud}.
In panel (a), the unshaded interior ellipse is the constraint from pion decay.
In the remaining plots, the allowed region is the band between the two ellipses.
An enlargement of the figures is displayed in figure \ref{fig11}.}
\label{bigi}
\end{figure}

\begin{figure}[ht!]
\newlength{\picwidthq}
\setlength{\picwidthq}{3in}
 \begin{center}
 \subfigure[][]{\resizebox{\picwidthq}{!}{\includegraphics{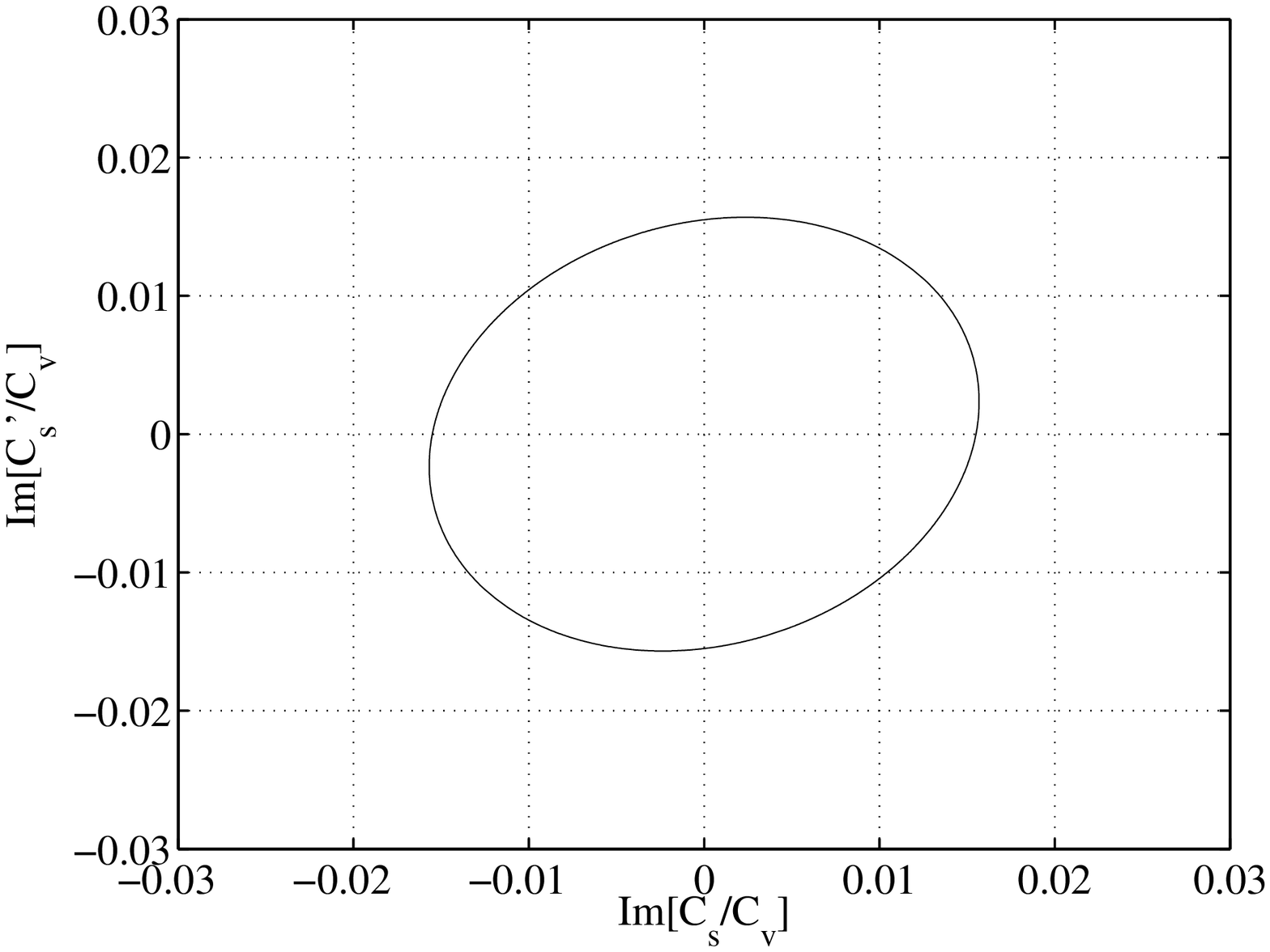}}}
 \subfigure[][]{\resizebox{\picwidthq}{!}{\includegraphics{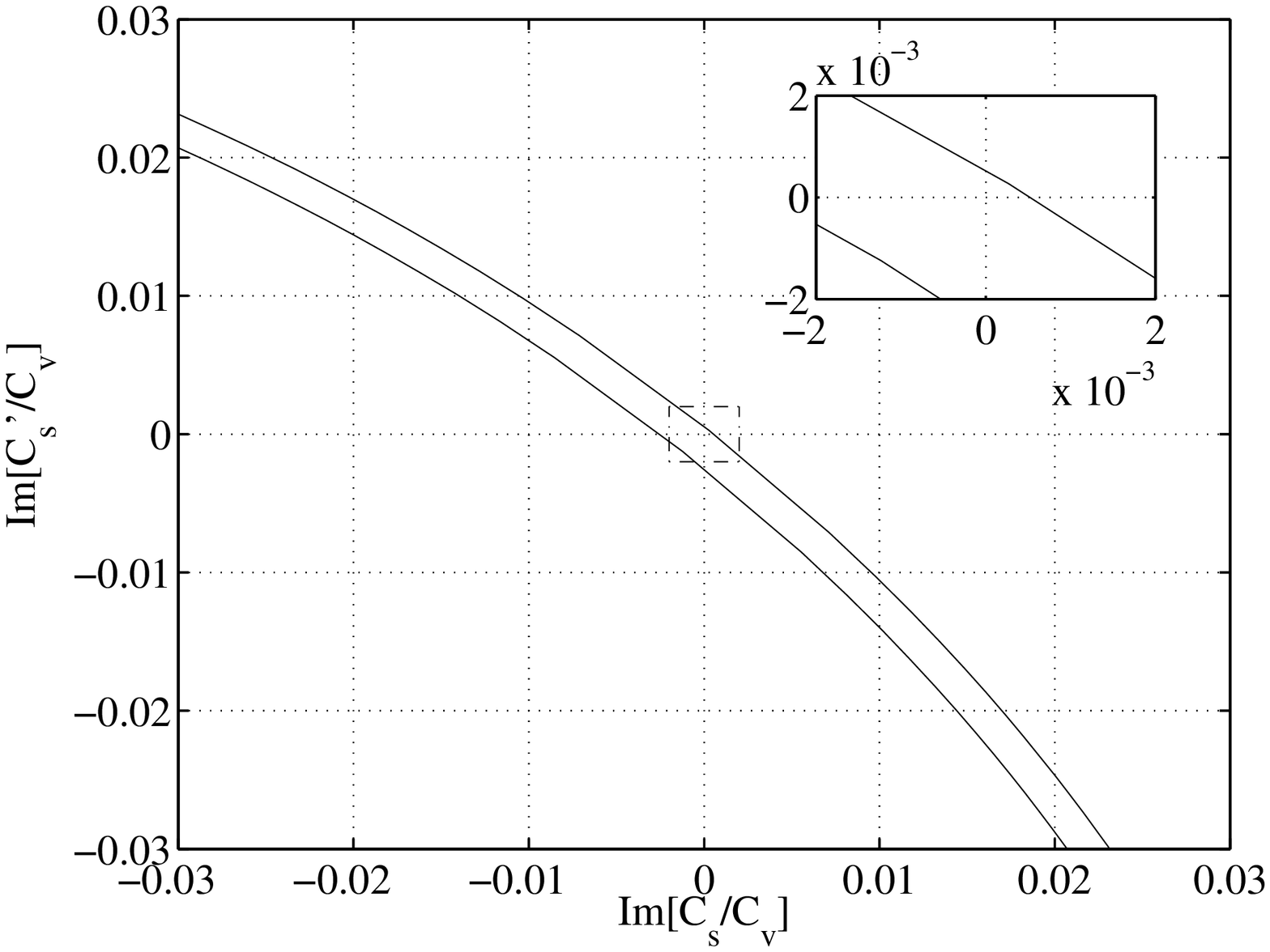}}}
 \subfigure[][]{\resizebox{\picwidthq}{!}{\includegraphics{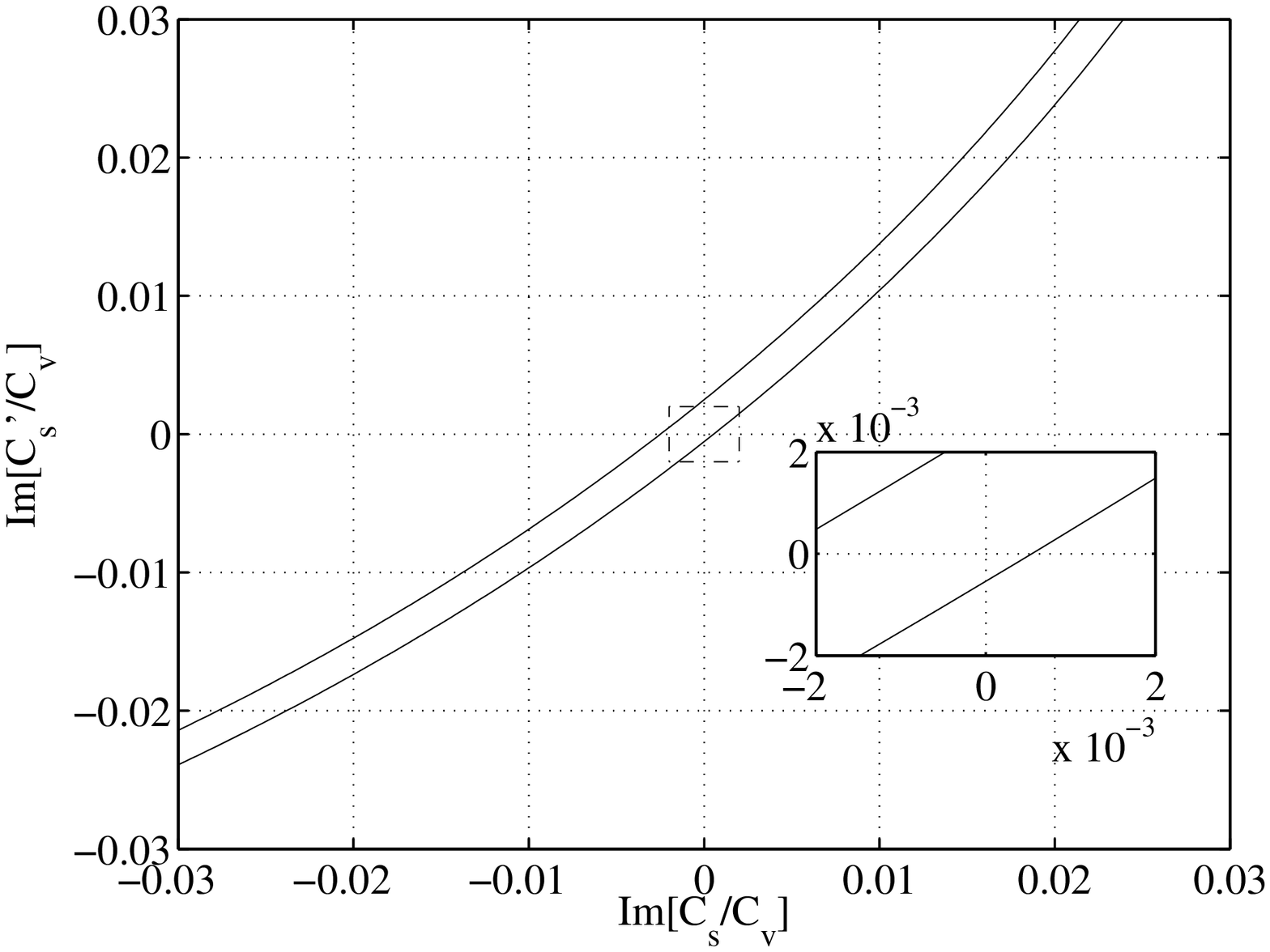}}}
 \end{center}
\caption{Constraint plots on the imaginary parts of
$\tilde C_s$ and $\tilde C_s^\prime$ at $\Lambda=$200 GeV. Panel (a)
corresponds to a phase of $\pm 90^\circ$; panel (b) to $\pm 45^\circ$;
and panel (c) to $45^\circ$ and $-45^\circ$ for $\tilde C_s$ and $\tilde C_s^\prime$ respectively.
In panel (a), the interior of the ellipse is the constraint.
In the remaining plots, the allowed region is the band between the two ellipses.
The enlarged area more clearly shows the width of the region.
}
\label{fig11}
\end{figure}

In the absence of right-handed neutrinos, if we consider $C_s$ and
$C_s^\prime$ to be purely real and the scale $\Lambda$ of the
order of 200 GeV, the indirect limits from $\pi^\pm \rightarrow
l^\pm\nu_l$ decay give us the limit
\be
-1.2 \times 10^{-3} \leq \tilde C_s \leq 2.7 \times 10^{-4}.
\ee
For comparison, the
experimental 90\% confidence limit determined from the b-Fierz
interference term in $\beta$-decay (see eq.(\ref{monster})) is
$|\mathrm{Re}(\tilde C_s)| \leq 8 \times 10^{-3}$
\cite{Adelberger:1999ud,Towner:1995za}. We see that the indirect
limit from pion decay is stronger by over an order of magnitude.
On the other hand, if we consider $C_s$ and $C_s^\prime$ to be
purely imaginary; again in the limit of left-handed couplings we
obtain,
\be
|\tilde C_s| \leq  1.2\times 10^{-2}
\ee
where the scale $\Lambda$ is of the order of 200 GeV.
Again for comparison, the experimental limit on the size of the
imaginary part at the 95\% confidence level, with only left-handed
neutrinos, is approximately $|\mathrm{Im}(\tilde C_s)| \leq 1
\times 10^{-1}$ \cite{Adelberger:1999ud}. The indirect $\pi^\pm
\rightarrow l^\pm\nu_l$ limit is stronger by approximately an order of
magnitude.
If we take  $\tilde C_s=-\tilde C_s^\prime$ so that we are in the limit of right-handed couplings and the b-Fierz interference term
vanishes we find,
\be
|\tilde C_s| \leq  1.0 \times 10^{-2}.
\ee
Again for comparison, at 1$\sigma$, the direct experimental constraint
is $|\tilde C_s| \leq 6 \times 10^{-2} \cite{Adelberger:1999ud}$.
In each case presented the scale of new physics was at $\Lambda =200$ GeV corresponding to
$\Delta_A(200 \hspace{1mm} \mathrm{GeV}) \approx 7.7 \times 10^{-4}$,
$\Delta_B(200 \hspace{1mm} \mathrm{GeV}) \approx 8.9 \times 10^{-4}$.
Because the pseudoscalar interactions are induced through renormalization group running
from $\Lambda$ down to the electroweak scale,
the higher the scale of new physics is, the more competitive our results become relative to beta decay.
As the scale of new physics is
lowered, the constraints from $\pi^\pm \rightarrow l^\pm\nu_l$ become less stringent. However even in the worst case limit
where the new scale is at the Z-mass and therefore we would no longer have an interval of renormalization group running,
the renormalization threshold effects calculated in eq.(\ref{thresres}) are still competitive.
As an example, if we take $C_s$ and $C_s^\prime$ to be real and ignore right-handed neutrinos we find that,
\be
-2 \times10^{-2} \leq \tilde C_s  \leq  4 \times 10^{-3}.
\label{effthres}
\ee

Plots of the pion physics constraints for the more general
situation (where the real and imaginary parts of $C_s$ and
$C_s^\prime$ vary independently) are given in figures \ref{bigr}, \ref{fig10}
\ref{bigi}, and \ref{fig11}. We plot the constraints for the real and
imaginary parts separately. Note from eq.(\ref{mstc}) that the
phases of $C_s$ and $C_s^\prime$ are important when constructing
these separate plots. In order to convey the effects of the phases
most clearly, we have chosen three interesting cases. In the real
plots we consider: $C_s$ and $C_s^\prime$ to each have a phase of
$0^\circ$; $C_s$ and $C_s^\prime$ to each have a phase of $\pm
45^\circ$; and the situation where $C_s$ has as a phase of
$45^\circ$ and $C_s^\prime$ has a phase of $-45^\circ$. In the
imaginary plots we consider: $C_s$ and $C_s^\prime$ to each have a
phase of $\pm 90^\circ$; $C_s$ and $C_s^\prime$ each have a phase
of $\pm 45^\circ$; and the case where $C_s$ has a phase of
$45^\circ$ and $C_s^\prime$ has a phase of $-45^\circ$. All three
plots in the imaginary case are well within the region allowed by
the direct experimental bounds
\cite{Adelberger:1999ud,Schneider:1983da}.

There are two points of interest that warrant further discussion.
First, note that in the limit of sufficiently large phases (i.e. $ >
85^\circ$) the ellipse bound in figure \ref{fig10} moves entirely
inside the b-Fierz interference limit allowed region. This is
expected since phases approaching $90^\circ$ imply that $C_s$ and
$C_s^\prime$ are almost completely imaginary. When this situation
occurs and we are in the limit of left-handed couplings (i.e.
along the line $C_s=C_s^\prime$), there are two solutions
consistent with the pion physics constraints and the b-Fierz
interference bound. One solution is centered around 0 and the other is
centered off 0 along the line $C_s=C_s^\prime$ yet inside the
b-Fierz interference limits. Even in these cases, the width of the
ellipse bound is still of the order of $2 \times 10^{-3}$.
Secondly, in order to move from the origin along the ellipse by more than
the width of the allowed region requires a delicate cancellation between the terms in
eq.(\ref{mstc}). If we ignore the possibility of this cancellation, the region allowed by
pion decay would collapse to a small region near the origin of length given by the width
of the ellipse bounds.

%%%%%%%%%%%%%%%%%%%%%%%%%%%%%%%%%%%%%%%%%%%%%%%%%%%%%

%%%%%%%%%%%%%%%%%%%%%%%%%%%%%%%%%%%%%%%%

\section{Flavour Dependent Couplings}
\label{lasts}
Thus far we have obtained limits on scalar interactions in the limit of universal flavour couplings.
Let us now relax this assumption. One case that deserves attention is the limit of mass proportional couplings.
This implies that $R_e/(m_e^2(m_\pi^2-m_e^2)) = R_\mu/(m_\mu^2(m_\pi^2-m_\mu^2))$ in eq.(\ref{Semu}) and therefore
there is no effect on the pion branching ratio,
\bea
\frac{\Gamma(\pi^- \rightarrow e \nu_e)}{\Gamma(\pi^- \rightarrow \mu \nu_\mu)}
&=&\frac{(m_\pi^2 -m_e^2)}{(m_\pi^2-m_\mu^2)} \left[\frac{m_e^2(m_\pi^2-m_e^2)
+S_e}{m_\mu^2(m_\pi^2-m_\mu^2) + S_\mu} \right] \nonumber \\
&=& T. \eea This observation also holds in the presence of
right-handed neutrinos. However, in this case, we still can bound
the scalar couplings involved in $\beta$-decay by combining the
$\pi^\pm \rightarrow l^\pm\nu_l$ limits with data from muon
capture experiments. Recent experiments and analysis of muon
capture on $^3\textnormal{He}$ indicate that the muon-nucleon
scalar coupling is bounded by \cite{Govaerts:2000ps} \be
\frac{|S_\mu|}{\Lambda^2} \leq 4 \times 10^{-2} \space G_F
\label{he3limit} \ee with a neutrino of left-handed chirality.
Therefore, in the limit of mass proportional couplings,
${S_e}/{\Lambda^2}$ must be of the order of 200 times smaller due
to the electron-muon mass ratio. This implies that $\tilde C_s$ is
bounded, \be |\tilde C_s| \leq 2 \times 10^{-4}. \label{mu} \ee

\begin{figure}[ht!]
\newlength{\picwidthk}
\setlength{\picwidthk}{3in}
 \begin{center}
\subfigure[][]{\resizebox{\picwidthk}{!}{\includegraphics{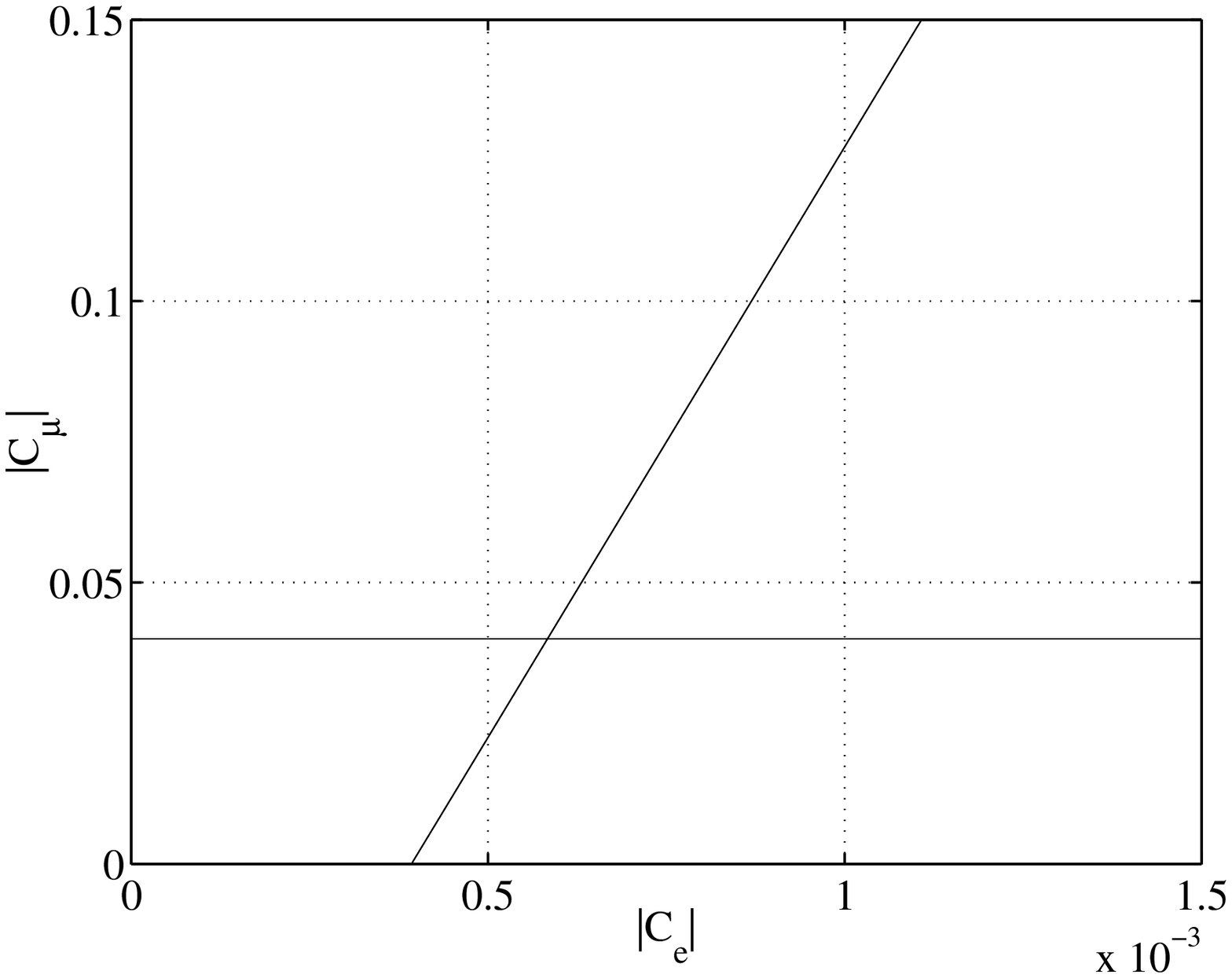}}}
 \subfigure[][]{\resizebox{\picwidthk}{!}{\includegraphics{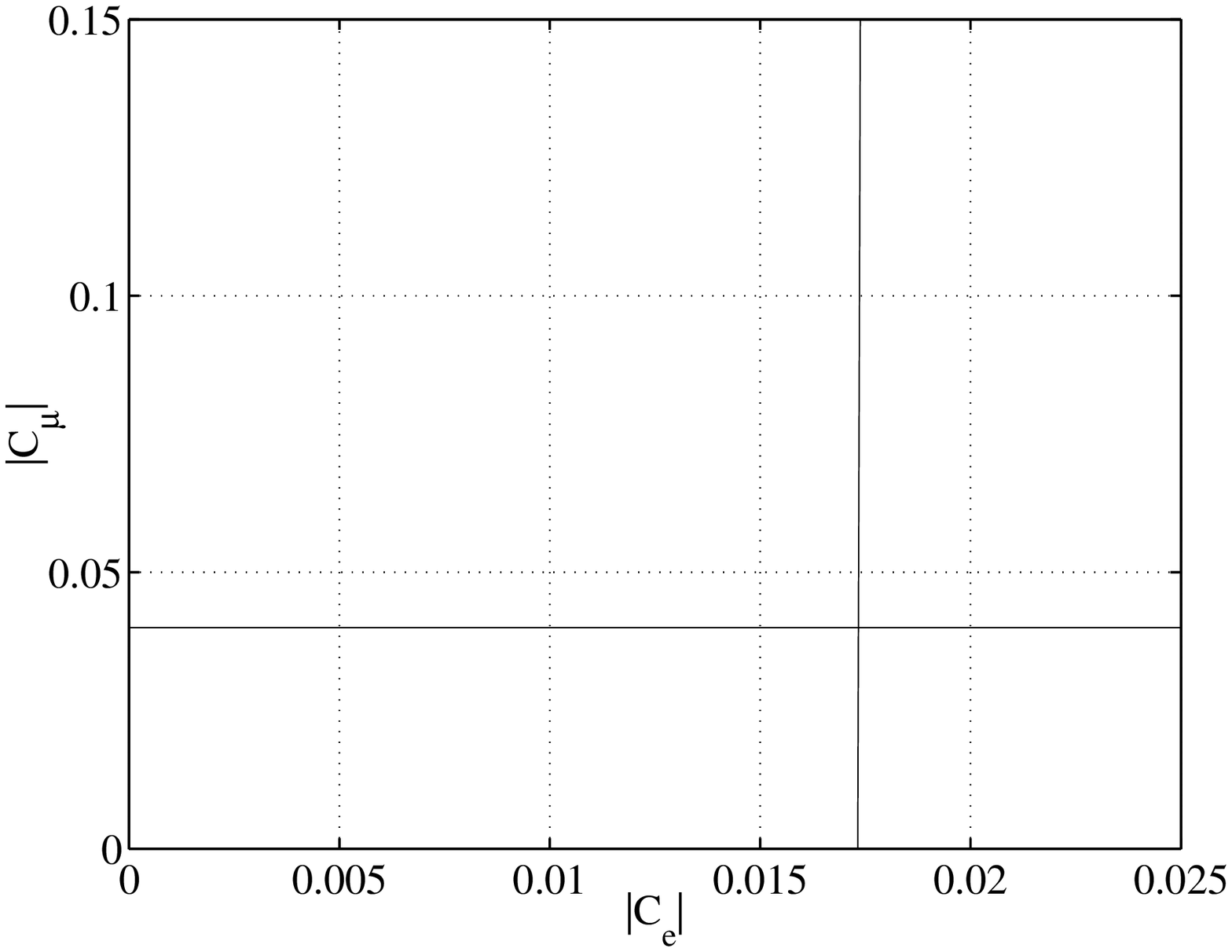}}}
 \subfigure[][]{\resizebox{\picwidthk}{!}{\includegraphics{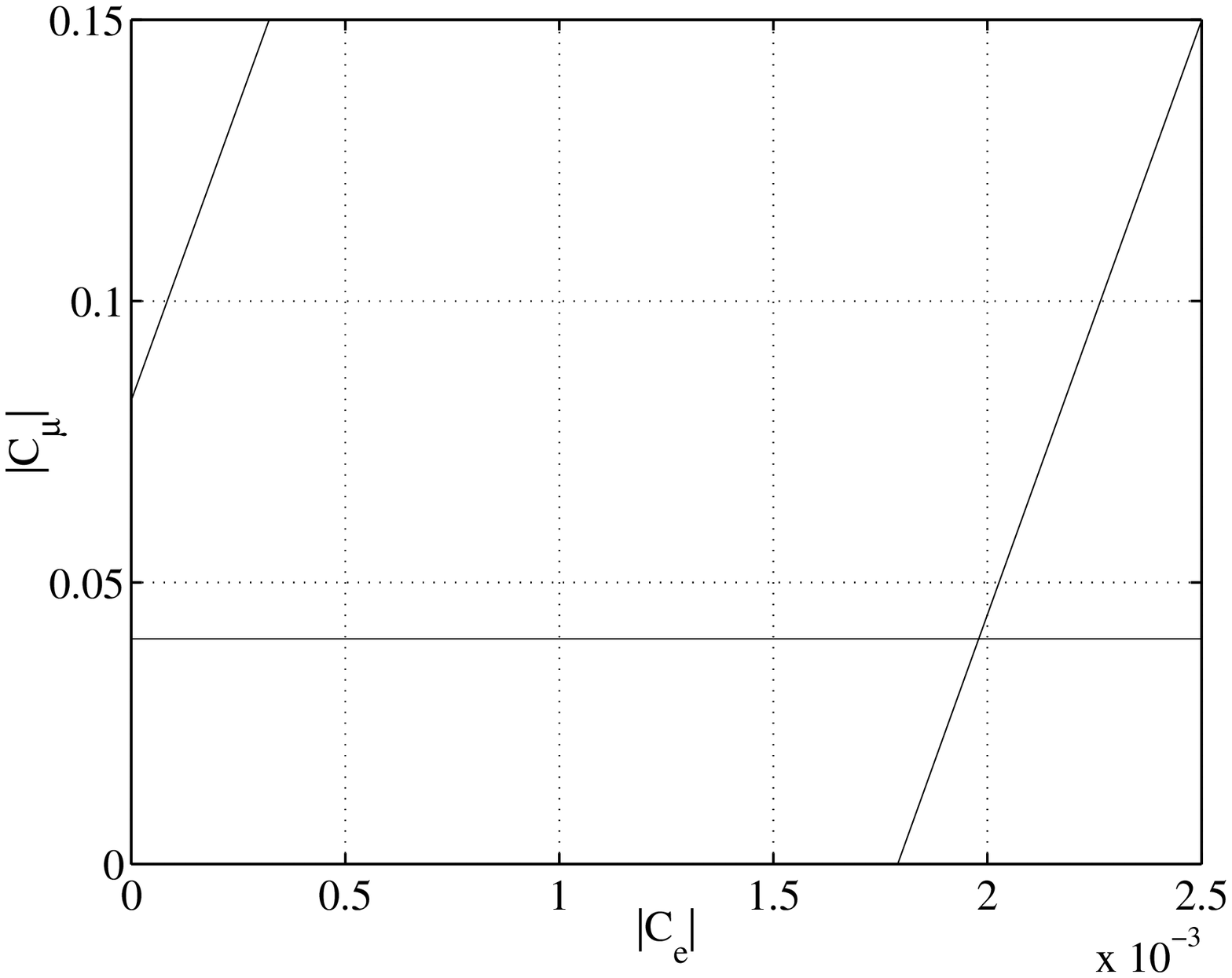}}}
 \subfigure[][]{\resizebox{\picwidthk}{!}{\includegraphics{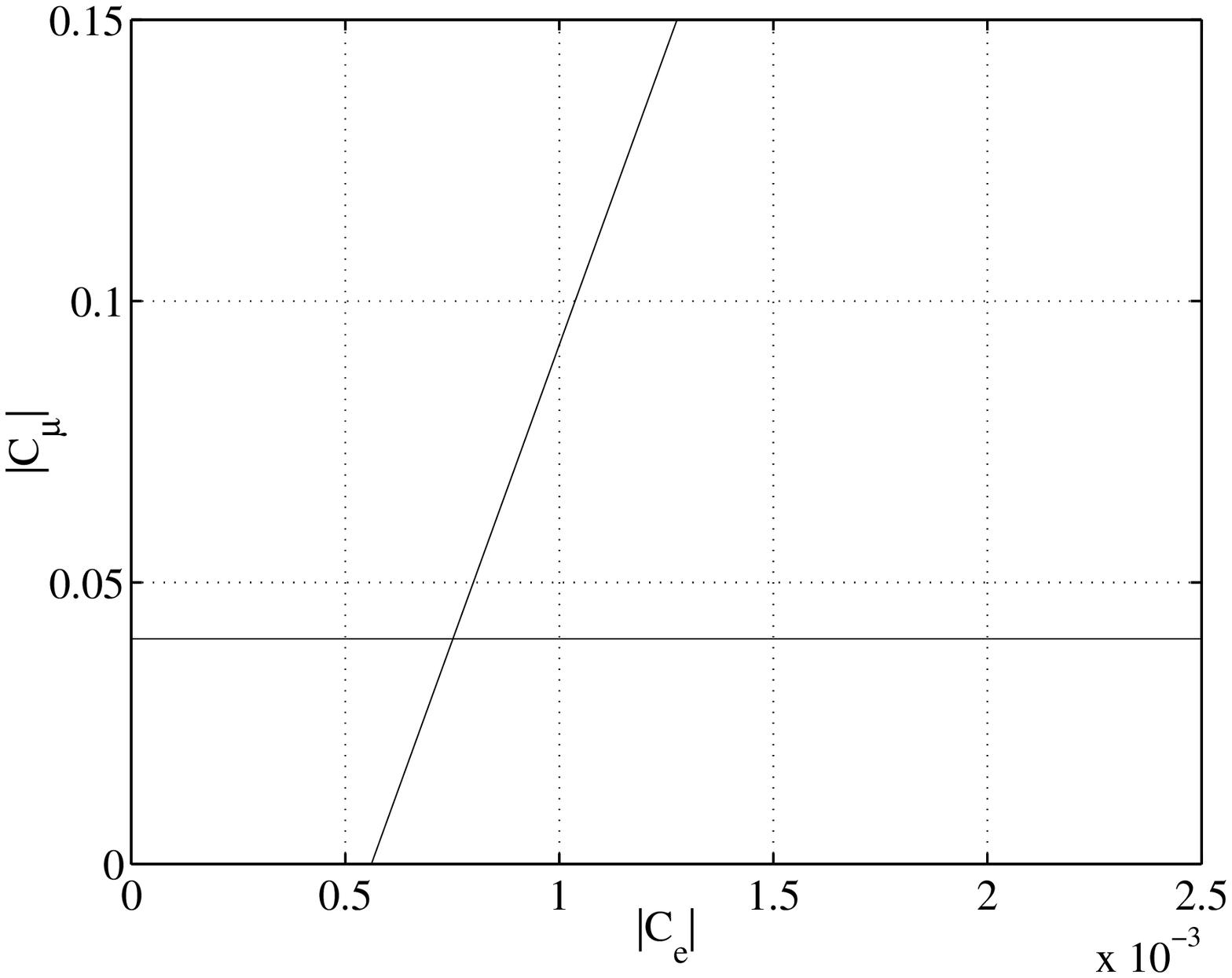}}}
 \end{center}
\caption{Constraint plots on the $|C_e|$ and $|C_\mu|$ couplings at $\Lambda = 200$ GeV.
Panel (a) corresponds to phases for $C_e$ and $C_\mu$ of $0^\circ$, $0^\circ$; panel (b) to $90^\circ$, $90^\circ$; panel (c) to
$180^\circ$, $180^\circ$; panel (d) to $\pm 45^\circ$, $\pm 45^\circ$ respectively.
The allowed region is the bounded area in the lower left corner. The horizontal line is the muon capture bound \cite{Govaerts:2000ps}.}
\label{emu}
\end{figure}

In order to estimate the degree to which the presence of muon scalar interactions can weaken the limits that we infer from
$\pi^\pm \rightarrow l^\pm \nu_l$, let us assume that the muon scalar coupling saturates the experimental bound eq.(\ref{he3limit}). Substituting
this into the expression for the pion branching ratio eq.(\ref{BRemu}), ignoring right-handed neutrinos and assuming a scale $\Lambda$ of 200 Gev,
eq.(\ref{csteq}) is modified to the following form,
\be
-3.3 \times 10^{-2}\leq \sqrt{2}\frac{\tilde f_\pi \mathrm{Re}(\rho)}{G_F \Lambda^2 f_\pi \cos \theta_c m_e}
+ \frac{|\rho|^2 \tilde f_\pi^2} {2 G_F^2 \Lambda^4 f_\pi^2 \cos^2 \theta_c  m_{e}^2}
 \leq 7.3 \space \times \space 10^{-3}.
\ee
We find this conservative approach has the effect of weakening our limits
a factor of three at most compared to the analysis in section \ref{betac}. The limits on scalar
couplings with $\Lambda = 200 \hspace{1mm} \mathrm{GeV}$, from $\pi^\pm \rightarrow l^\pm\nu_l$ combined with muon capture
scalar limits, are substantially stronger than limits on scalar couplings from direct $\beta$-decay searches.

Finally, we consider the allowed region for the electron-scalar and muon-scalar couplings in a model independent manner.
Again the constraint equation derived from eq.(\ref{BRemu}) is,
\be
-1.0\times 10^{-2} \leq \left(\frac{1 +
\sqrt{2}\frac{\tilde f_\pi \mathrm{Re}(C_e) \Delta_A}{f_\pi \cos \theta_c m_e} + \frac{|C_e|^2 \Delta_A^2 \tilde f_\pi^2}
{2 f_\pi^2 \cos^2 \theta_c m_{e}^2}}{1 +
\sqrt{2}\frac{\tilde f_\pi \mathrm{Re}(C_\mu) \Delta_A}{f_\pi \cos \theta_c m_\mu} + \frac{|C_\mu|^2 \Delta_A^2
\tilde f_\pi^2}{2 f_\pi^2 \cos^2 \theta_c m_{\mu}^2}} - 1  \right)  \leq  2.2 \times 10^{-3},
\ee
where,
\be
C_\mu = \frac{S_\mu}{G_F \Lambda^2} \hspace{5mm} C_e = \frac{S_e}{G_F \Lambda^2}.
\ee
We display the results in figure \ref{emu} for a number of different phase conditions.
We consider the cases where the complex phase of $C_e$ and $C_\mu$ are $0^\circ$, $0^\circ$;
$90^\circ$, $90^\circ$; $180^\circ$, $180^\circ$; $\pm 45^\circ$, $\pm 45^\circ$, respectively.

\section{Discussion}

By considering renormalization effects on universal (or
alternatively first generation), and flavour diagonal scalar
operators, we have derived limits on the size of the ratio between
scalar and vector couplings from precision measurements of
$\pi^\pm \rightarrow l^\pm\nu_l$ decay. As a typical constraint
value, in the absence right-handed neutrinos, we find that
$-1.2 \times 10^{-3} \leq \tilde C_s \leq 2.7 \times 10^{-4}$ for $\Lambda$ of
the order of 200 GeV. A more general comparison with the
$\beta$-decay experiments (with the inclusion of right-handed
neutrinos) is made in the plots in figure \ref{fig10} and figure
\ref{fig11}. We note that the most conservative estimate of the
limits occurs when the new physics arises at the electroweak
scale. In this case, the contribution to the induced pseudoscalar
comes entirely from threshold corrections which we estimate from
the calculations in section \ref{modelcal}. The limit for real couplings
in the absence of right-handed neutrinos from threshold
contributions is $-3 \times10^{-2} \leq \tilde C_s  \leq  6\times 10^{-3}$.
In the scenario where we have arbitrary generation dependence of
the scalar couplings, $\pi^\pm \rightarrow l^\pm\nu_l$ limits can
be combined with limits on scalar interactions in muon capture to
bound the first generation scalar couplings. These limits are
illustrated in particular cases in figure \ref{emu}.

These observations have implications for current $\beta$-decay experiments. Direct searches for scalar interactions in $\beta$-decay will be
most competitive if the new physics responsible for the effective scalar interactions
arises at the electroweak scale in the explicit exchange of new scalar particles. In these circumstances, the indirect
limits from threshold induced pseudoscalar interactions, eq.(\ref{effthres}), are comparable to the direct $\beta$-decay scalar searches.
Therefore, interest in searches for new scalar interactions with $\beta$-decay experiments remains undiminished.

On the other hand, for new effective scalar interactions arising as effective SU(2) $\times$ U(1) invariant operators at mass scales above 200 GeV (as
expected in models with leptoquarks, composite quarks/leptons, or low scale quantum gravity) the constraints arising from the precision measurements of
$\pi^\pm \rightarrow l^\pm\nu_l$ decay, combined with limits on scalar interactions in muon capture, can be stronger by an order of magnitude or more
than the direct experimental searches. Furthermore, the relative strength of these searches becomes better, the higher the mass scale of the new physics
compared to the electroweak scale. This argues strongly for improved experimental precision in measurements of muon capture, and
$\pi^\pm \rightarrow l^\pm\nu_l$ decay. In particular we note that in the case of pion decay, the experimental error exceeds the uncertainty in the
theoretical calculation by a factor of eight. A new measurement of $\pi^\pm \rightarrow l^\pm\nu_l$ decay with an order of magnitude greater precision would
not only constrain physics beyond the standard model which could potentially contribute to tree level pion decay, but as we have argued above,
will also indirectly provide tests of new scalar interactions of unparalleled precision.

\section{Acknowledgements}
We would like to thank Eric Adelberger and John Behr for their generous assistance at several stages of this project.
We are especially grateful for the help that John Behr and Manuella Vincter extended to us
in regards to issues in error analysis.
We have benefitted from useful discussions with Doug Bryman, Andrzej Czarnecki, John Ellis, Randy Lewis, Michael Luke, and Aneesh Manohar.
We wish to thank J.P. Archambault, Eric Carpenter and David Shaw
for their help with the Feynmf package and graphics. This work was supported by the Natural Sciences and Engineering Research Council of Canada.


\begin{thebibliography}{10}
\providecommand*{\bibinfo}[2]{#2}
\providecommand*{\eprint}[1]{#1}
\providecommand*{\url}[1]{#1}
\bibitem{PDG}
\bibinfo{author}{Particle Data Group}, \bibinfo{journal}{Phys. Rev.}
  \bibinfo{volume}{\textbf{D66}},(\bibinfo{date}{2002})
\bibitem{Garcia:1999da}
 \bibinfo{author}{A.~Garcia} \emph{et~al.}, \eprint{nucl-ex/9904001}.
\bibitem{Adelberger:1999ud}
\bibinfo{author}{E.~G. Adelberger} \emph{et~al.}
  (\bibinfo{collaboration}{ISOLDE}), \bibinfo{journal}{Phys. Rev. Lett.}
  \bibinfo{volume}{\textbf{83}}, \bibinfo{pages}{1299} (\bibinfo{date}{1999}),
  \eprint{nucl-ex/9903002}.
\bibitem{Adelberger:1993wq}
\bibinfo{author}{E.~G. Adelberger}, \bibinfo{journal}{Phys. Rev. Lett.}
  \bibinfo{volume}{\textbf{70}}, \bibinfo{pages}{2856} (\bibinfo{date}{1993}).
\bibitem{Towner:1995za}
\bibinfo{author}{For a review see: I.~S. Towner} and \bibinfo{author}{J.~C. Hardy}, \eprint{nucl-th/9504015}.
\bibitem{Bryman}
\bibinfo{author}{For a review see: D.~A. Bryman} \bibinfo{journal}{Comments Nucl. Part. Phys.} \bibinfo{volume}{\textbf{21}}
\bibinfo{pages}{101} (\bibinfo{date}{1993}).
\bibitem{Marciano:1993sh}
\bibinfo{author}{W.~J. Marciano} and \bibinfo{author}{A.~Sirlin},
\bibinfo{journal}{Phys. Rev. Lett.} \bibinfo{volume}{\textbf{71}},
\bibinfo{pages}{3629} (\bibinfo{date}{1993}).
\bibitem{Britton1}
 \bibinfo{author}{D.~I. Britton} \emph{et~al.},\bibinfo{journal}{Phys. Rev. Lett.}
  \bibinfo{volume}{\textbf{68}}, \bibinfo{pages}{3000-3003}
  (\bibinfo{date}{1992}).
\bibitem{Britton2}
 \bibinfo{author}{D.~I. Britton} \emph{et~al.},\bibinfo{journal}{Phys. Rev. D}
  \bibinfo{volume}{\textbf{49}}, \bibinfo{pages}{17-20}
  (\bibinfo{date}{1993}).
\bibitem{Czapek1}
 \bibinfo{author}{D.~I. Britton} \emph{et~al.},\bibinfo{journal}{Phys. Rev. Lett.}
  \bibinfo{volume}{\textbf{70}}, \bibinfo{pages}{3000-3003}
  (\bibinfo{date}{1992}).
\bibitem{Manohar:1996cq}
For reviews see: \\
 \bibinfo{author}{A.~Manohar}, \eprint{hep-ph/9606222} \\
 \bibinfo{author}{D.~Kaplan}, \eprint{nucl-th/9506035}.
\bibitem{Weinberg:1979}
\bibinfo{author}{S.~Weinberg}, \bibinfo{journal}{Physica}
\bibinfo{volume}{\textbf{96A}}, \bibinfo{pages}{327-340} (\bibinfo{date}{1979}).
\bibitem{Wright1}
\bibinfo{author}{H.~Arason}, \bibinfo{author}{D.~J.~ Castano}, \bibinfo{author}{B.~Keszthelyi},
\bibinfo{author}{S.~Mikaelian}, \bibinfo{author}{E.~J.~Piard}, \bibinfo{author}{P.~Ramond}, and
\bibinfo{author}{B.~D.~Wright}
\bibinfo{journal}{Phys.Rev.} \bibinfo{volume}{\textbf{D46}}, \bibinfo{pages}{3945-3965} (\bibinfo{date}{1992}).
\bibitem{Wright2}
\bibinfo{author}{B.~D.~Wright}
\eprint{hep-ph/9404217}
\bibitem{Jack:1957}
\bibinfo{author}{J.~Jackson}, \bibinfo{author}{S.~Treiman}, and
\bibinfo{author}{H.~Wyld~Jr.}, \bibinfo{journal}{Nucl. Phys.}
\bibinfo{volume}{\textbf{4}}, \bibinfo{pages}{206} (\bibinfo{date}{1957}).
\bibitem{Liu:1998um}
\bibinfo{author}{S.~J. Dong, J.~F. Laga\"e}, and \bibinfo{author}{K.~F. Liu}, \bibinfo{journal}{Phys. Rev.}
\bibinfo{volume}{\textbf{D54}}, \bibinfo{pages}{5496-5500}
(\bibinfo{date}{1996}), \eprint{hep-ph/9602259}.
\bibitem{Schneider:1983da}
\bibinfo{author}{M.~B. Schneider} \emph{et~al.}, \bibinfo{journal}{Phys. Rev.
Lett.} \bibinfo{volume}{\textbf{51}},
\bibinfo{pages}{1239} (\bibinfo{date}{1983}).
\bibitem{Govaerts:2000ps}
\bibinfo{author}{J.~Govaerts} and \bibinfo{author}{J.-L. Lucio-Martinez},
\bibinfo{journal}{Nucl. Phys.} \bibinfo{volume}{\textbf{A678}},
\bibinfo{pages}{110} (\bibinfo{date}{2000}), \eprint{nucl-th/0004056}.


\end{thebibliography}
\end{document}